\documentclass[useAMS, fleqn, usenatbib]{mnras}

\usepackage{amsmath}
\usepackage[T1]{fontenc}
\usepackage{ae,aecompl}

\usepackage{lmodern}
\usepackage{graphicx}
\usepackage[flushleft]{threeparttable}
\newcommand\ltsima{$\; \buildrel < \over \sim \;$}
\newcommand\simlt{\lower.5ex\hbox{\ltsima}}
\newcommand\gtsima{$\; \buildrel > \over \sim \;$}
\newcommand\simgt{\lower.5ex\hbox{\gtsima}}

  \newcommand{\bc}{\begin{center}}
  \newcommand{\ec}{\end{center}}
  
  \newcommand{\hMsun}{~{\rm h^{-1}}\>{\rm M_{\sun}}}
  \newcommand{\mpc}{~{\rm h^{-1}}~{\rm Mpc}}
  \newcommand{\kpc}{~{\rm h^{-1}}~{\rm kpc}}

\newcommand{\art}{\textsc{Art}}

\newcommand{\arepo}{\textsc{Arepo}}
\newcommand{\hydra}{\textsc{Hydra}}

\newcommand{\gadget}{\textsc{Gadget}}
\newcommand{\gadgetthree}{\textsc{Gadget-3}}
\newcommand{\gadgetx}{\textsc{G3-x}}
\newcommand{\gadgetmagneticum}{\textsc{G3-Magneticum}}
\newcommand{\gadgettwox}{\textsc{G2-x}}
\newcommand{\gadgetowls}{\textsc{G3-Owls}}

\newcommand{\gadgetmusic}{\textsc{G3-Music}}
\newcommand{\gadgetsphs}{\textsc{G3-Sphs}}
\newcommand{\gadgetpesph}{\textsc{G3-Pesph}}
\newcommand{\gadgetanarchy}{\textsc{G2-Anarchy}}
\newcommand{\ramses}{\textsc{Ramses}}

\newcommand{\arepoil}{\textsc{Arepo-IL}}
\newcommand{\areposh}{\textsc{Arepo-SH}}

\newcommand{\gadgetmusicpi}{\textsc{G2-Musicpi}}

\defcitealias{Sembolini2015}{Paper~I}
\defcitealias{Sembolini2015b}{Paper~II}
\defcitealias{Elahi2015}{Paper~III}

\title[nIFTy IV: the influence of baryons]{nIFTy Galaxy Cluster
  simulations IV: Quantifying the Influence of Baryons on Halo Properties}

\author[Weiguang Cui et al.]
{\parbox{\textwidth}{Weiguang Cui,$^{1,2}$\thanks{E-mail: \texttt{weiguang.cui@uwa.edu.au}}
    Chris Power,$^{1,2}$ Alexander Knebe,$^{3,4}$ Scott T. Kay,$^5$
    Federico Sembolini,$^{3,4}$ Pascal J. Elahi,$^6$ Gustavo Yepes,$^{3,4}$
    Frazer Pearce,$^7$ Daniel Cunnama,$^8$ Alexander M. Beck,$^{9}$ Claudio Dalla Vecchia,$^{10,11}$
    Romeel Dav\'e,$^{12,13,14}$ Sean February,$^{15}$ Shuiyao Huang,$^{16}$
    Alex Hobbs,$^{17}$ Neal Katz,$^{16}$ Ian G. McCarthy,$^{18}$
    Giuseppe Murante,$^{19}$ Valentin Perret,$^{20}$ Ewald Puchwein,$^{21}$
    Justin I. Read,$^{22}$ Alexandro Saro,$^{9,23}$
    Romain Teyssier,$^{23}$ and Robert J. Thacker$^{24}$}\vspace{0.4cm}\\
  \parbox{\textwidth}{
    $^1$ ICRAR, University of Western Australia, 35 Stirling Highway,
    Crawley, Western Australia 6009, Australia\\
    $^2$ ARC Centre of Excellence for All-Sky Astrophysics (CAASTRO)\\
    $^3$ Departamento de F\'isica Te\'orica, M\'odulo 8, Facultad de Ciencias,
    Universidad Aut\'onoma de Madrid, 28049 Madrid, Spain\\
    $^4$ Astro-UAM, UAM, Unidad Asociada CSIC\\
    $^6$ Sydney Institute for Astronomy, University of Sydney, Sydney, NSW 2016,
    Australia\\
    $^7$ School of Physics \& Astronomy, University of Nottingham, Nottingham
    NG7 2RD, UK\\
    $^5$ Jodrell Bank Centre for Astrophysics, School of Physics and Astronomy,
    The University of Manchester, Manchester M13 9PL, UK \\
    $^8$ Physics Department, University of the Western Cape, Cape Town 7535,
    South Africa\\
    $^9$ University Observatory Munich, Scheinerstr. 1, D-81679 Munich, Germany\\
    $^{10}$ Instituto de Astrof\'isica de Canarias, C/V\'ia L\'actea s/n, 38205 La Laguna, Tenerife, Spain\\
    $^{11}$ Departamento de Astrof\'isica, Universidad de La Laguna, Av. del Astrof\'isico Francisco S\'anchez s/n, 38206 La Laguna, Tenerife, Spain\\
    $^{12}$ Physics Department, University of Western Cape, Bellville, Cape Town
    7535, South Africa\\
    $^{13}$ South African Astronomical Observatory, PO Box 9, Observatory, Cape
    Town 7935, South Africa\\
    $^{14}$ African Institute of Mathematical Sciences, Muizenberg, Cape Town
    7945, South Africa\\
    $^{15}$ Center for High Performance Computing, CSIR Campus, 15 Lower Hope
    Street, Rosebank, Cape Town 7701, South Africa\\
    $^{16}$ Astronomy Department, University of Massachusetts, Amherst, MA
    01003, USA\\
    $^{17}$ Institute for Astronomy, Department of Physics, ETH Zurich,
    Wolfgang-Pauli-Strasse 16, CH-8093, Zurich, Switzerland\\
    $^{18}$ Astrophysics Research Institute, Liverpool John Moores University,
    146 Brownlow Hill, Liverpool L3 5RF, UK\\
    $^{19}$ INAF - Osservatorio Astronomico di Trieste, via G.B. Tiepolo 11,
    I-34143 Trieste, Italy\\
    $^{20}$ Centre for Theoretical Astrophysics and Cosmology, Institute for
    Computational Science, University of Zurich, Winterthurerstrasse 190, 8057
    Zurich, Switzerland\\
    $^{21}$ Institute of Astronomy and Kavli Institute for Cosmology, University
    of Cambridge, Madingley Road, Cambridge CB3 0HA, UK\\
    $^{22}$ Department of Physics, University of Surrey, Guildford, GU2 7XH,
    Surrey, United Kingdom\\
    $^{23}$ Excellence Cluster Universe, Boltzmannstr. 2, 85748 Garching, Germany\\
    $^{24}$ Department of Astronomy and Physics, Saint Mary's University, 923
    Robie Street, Halifax, Nova Scotia, B3H 3C3, Canada\\
  }
}

\date{Accepted XXX. Received YYY; in original form ZZZ}
\pubyear{2015}

\begin{document}
\label{firstpage}
\pagerange{\pageref{firstpage}--\pageref{lastpage}}
\maketitle

\clearpage

\begin{abstract}

  Building on the initial results of the nIFTy simulated galaxy cluster
  comparison, we compare and contrast the impact of baryonic physics with a single massive galaxy cluster, run with 11 state-of-the-art codes, spanning adaptive mesh, moving mesh, classic and modern SPH approaches. For each code represented we have a dark matter only (DM) and non-radiative (NR) version of the cluster, as well as a full physics (FP) version for a subset of the codes.
  We compare both radial mass and kinematic profiles, as well as global measures
  of the cluster (e.g. concentration, spin, shape), in the NR and
  FP runs with that in the DM runs. Our analysis reveals good consistency ($\loa 20 \%$) between global properties of the cluster predicted by different codes when integrated quantities are measured within the virial radius $R_{200}$. However, we see larger differences for quantities within $R_{2500}$, especially in the FP runs. The radial profiles reveal a diversity, especially in the cluster centre, between the NR runs, which can be understood straightforwardly from the division of codes into classic SPH and non-classic SPH (including the modern SPH, adaptive and moving mesh codes); and between the FP runs, which can also be understood broadly from the division of codes into those that include AGN feedback and those that do not. The variation with respect to the median is much larger in the FP runs with different baryonic physics prescriptions than in the NR runs with different hydrodynamics solvers.

\end{abstract}
\begin{keywords}
methods: N-body simulations -- galaxies: clusters: general -- galaxies: haloes
-- galaxies: evolution -- cosmology: theory -- galaxies: formation
\end{keywords}

%*****************************************************************************

\section{Introduction}
\label{sec:introduction}
The importance of galaxy clusters as probes of cosmology and testbeds for galaxy
transformation and evolution is well recognised
\citep[e.g.][]{KravtsovBorgani2012}. Numerical simulations are fundamental to
give an accurate interpretation of the astrophysical processes observed in
galaxy clusters \citep[e.g.][]{Borgani2011}. %Accurate interpretation of the
%astrophysical significance of observations of galaxy clusters relies upon the
%predictions of numerical simulations.
Cosmological $N$-body simulations have
been used to estimate the abundance of galaxy clusters as a function of
redshift, which can be used to constrain values of the cosmological parameters
such as $\sigma_8$ \citep[e.g.][]{Viel2006} and the dark energy equation of
state \citep[e.g.][]{Angulo2005}, and to calibrate observational estimators of
cluster mass \citep[e.g.][]{Fabjan2011,Kay2012,Munari2013} and sensitivity to
dynamical state \citep[e.g.][]{Power2012}.

Cosmological hydrodynamical simulations offer the potential to test galaxy
transformation within cluster environments, although this has proven to be more
challenging. The Santa Barbara Cluster Comparison \citep{Frenk1999} already
highlighted that simulations of the same object performed with different codes can produce divergent
behaviour, most compactly quantified by the spherically averaged entropy profile
-- Eulerian mesh-based codes predicted entropy cores while Lagrangian SPH codes
predicted continuously declining entropy with decreasing radius. Subsequent
studies demonstrated that this divergent behaviour could be traced to the
treatment of surface tension and the suppression of multi-phase fluid mixing in
the classic SPH codes \citep[e.g.][]{Wadsley2008,Mitchell2009,Power2014,Sembolini2015}.

Given the developments in astrophysical simulation codes, as well as the
implementations of the hydrodynamic evolution of the baryons, after
$\sim$ 15 years of the Santa Barbara Cluster Comparison, it was natural to
investigate how the state-of-the-art codes compared when faced with the same
problem -- that of the
thermodynamical structure of a massive galaxy cluster at $z$=0, when only
gravity and non-radiative hydrodynamics is modelled. This formed the basis of
the nIFTy galaxy cluster comparison, the first results of which were presented
in \citet[][hereafter Paper~I]{Sembolini2015}. Initially, thirteen different
codes -- \art, \arepo, \hydra, \ramses\ and 9 incarnations of \gadget\ -- were
used to simulate a massive galaxy cluster down to $z$=0.
The mesh-based codes \art\ and \arepo\ formed extended entropy cores in the gas with
rising central gas temperatures, whereas ``classic'' SPH codes produced falling
entropy profiles all the way into the very centre with correspondingly rising
mass profiles and central temperature inversions. In contrast, modern SPH
codes produce gas entropy profiles that are essentially indistinguishable
from those obtained with mesh-based codes.

Building on the work presented in \citetalias{Sembolini2015}, \citet[][hereafter Paper~II]{Sembolini2015b}
compared these codes with different
radiative physical implementations -- such as cooling, star formation and AGN
feedback -- and showed that adding radiative physics washes away the
marked code-based differences present in the entropy profile seen in the
non-radiative simulations presented in \citetalias{Sembolini2015}.

\citet[][hereafter Paper~III]{Elahi2015} found that subhalo properties are reasonably consistent across almost all codes in DM, NR and FP simulations, although the code-to-code scatter increases with the inclusion of gas and subgrid baryonic physics. In the FP runs, the synthetic galaxies that reside in these subhaloes show striking code-to-code variation, with differences in stellar and gas masses being up to 0.2-0.4 dex.

In this paper, we follow up on the results presented
in \citetalias{Sembolini2015,Sembolini2015b,Elahi2015}, and focus on
how the inclusion of the baryonic component modifies the spatial and
kinematic structure of the simulated cluster. % in the $N$-body only limit.
We seek to understand
\begin{itemize}
  \item {\it the scatter between simulation codes and different
    input baryon models; and }
  \item {\it the effects of input baryon models on cluster properties,
    as well as the extent to which they converge.}
\end{itemize}
We consider the global properties of the cluster -- concentration, spin
parameter, inner slope, masses, halo shapes, and velocity anisotropy. The
cluster mass is calculated within the radii containing overdensities of 200,
500 and 2500 times the critical density of the Universe at $z$=0 (i.e. $R_{200}$,
$R_{500}$, $R_{2500}$). Halo shapes, as measured for isodensity and isopotential
surfaces, and velocity anisotropy are calculated
at these three radii. We also investigate the density, circular velocity and
velocity dispersion profiles.

The paper is organised as follows. In \S\ref{sec:simulation_codes}, we provide
a brief summary of the main features of the astrophysical simulation codes used
in this study, while in \S\ref{sec:simulated_cluster}, we recall the key
properties of the simulated galaxy cluster used in the comparison. The main
results are presented in \S\ref{sec:pro} and \S\ref{sec:gp}, in which we
investigate how the presence of a non-radiative and radiative physical baryonic
influences the simulated cluster. Finally in \S\ref{sec:conclusions}, we
discuss our results and state our conclusions.

%%%%%%%%%%%%%%%%%%%%%%%%%%%%%%%%%%%%%%%%%%%%%%%%%%%
\section{The Simulation Codes} \label{sec:simulation_codes}
%%%%%%%%%%%%%%%%%%%%%%%%%%%%%%%%%%%%%%%%%%%%%%%%%%%

\begin{table*}
  \caption{Brief summary of all the simulation codes participating in the nIFTy
    cluster comparison project.}
  \label{tab:codes}
  \begin{threeparttable}
  \begin{center}
    \begin{tabular}{llllll}
      \hline
    Type &  Code name, Reference  & & Baryonic models & & \\
      \hline
      & & DM & NR  & FP \\
      & & gravity solver & gas treatment & noAGN & AGN \\
      \hline \hline
      Grid-based & \ramses,       \cite{Teyssier2002} & AMR & \vtop{\hbox{ Godunov scheme}\vspace{1pt}\hbox{ with Riemann solver}} & N & Y \\
      \\
      Moving-mesh & \arepo,      \cite{Springel10} & TreePM & \vtop{\hbox{ Godunov scheme }\vspace{1pt}\hbox{ on moving mesh}} & Y\tnote{a} & Y\tnote{b} \\
      \\
     & \gadgetanarchy,  Dalla Vecchia et al. {\rm in prep.} & TreePM & SPH
     kernel: Wendland C2 & N & N\\
     & \gadgetsphs,   \cite{Read2012} & TreePM & Wendland C4 & N & N\\
     Modern SPH & \gadgetmagneticum,   \cite{Magn14} & TreePM & Wendland C6
     & N & Y\\
     & \gadgetx,  \cite{Beck2016} & TreePM & Wendland C4 & N & Y\\
     & \gadgetpesph, Huang et al. {\rm in prep.} & TreePM & HOCTS B-spline & Y & N\\
      \\
    &  \gadgetmusic,  \cite{Sembolini2013} & TreePM & Cubic spline & Y\tnote{c} & N\\
    Classic SPH &  \gadgetowls,  \cite{Schaye10} & TreePM & Cubic spline & N & Y\\
    &  \gadgettwox,  \cite{Pike2014} & TreePM & Cubic spline & N & Y\\
    &  \hydra,	  \cite{Couchman95} & {$\rm AP^3M$} & Cubic spline & N & N\\
      \hline
    \end{tabular}
    \begin{tablenotes}
    \item[a] This version is named \areposh.
    \item[b] This version is named \arepoil.
    \item[c] Two versions (\gadgetmusic\ and \gadgetmusicpi) are included in this model.
    \end{tablenotes}
  \end{center}
  \end{threeparttable}
\end{table*}

Following the classification adopted in nIFTy \citetalias{Sembolini2015, Sembolini2015b}, the 11 simulation
codes used in this study are divided into four groups based on their gas
dynamic solving techniques:
\begin{itemize}
\item Grid-based: -- \ramses\ \citep{Teyssier2002};
\item Moving-mesh: -- \arepo\ \citep{Springel10};
\item Modern SPH: -- \gadgetanarchy\ (Dalla Vecchia et al. {\rm in prep.}),
  \gadgetsphs\ \citep{Read2012}, \gadgetmagneticum\ \citep{Magn14},
  \gadgetx\ \citep{Beck2016}, \gadgetpesph\ (Huang et al. {\rm in prep.}); and
\item Classic SPH: -- \gadgetmusic\ \citep{Sembolini2013}, \gadgetowls\
  \citep{Schaye10}, \gadgettwox\ \citep{Pike2014}, \hydra\ \citep{Couchman95}.
\end{itemize}
For each simulation code we have dark-matter-only (DM) runs and non-radiative
(NR) runs, which include both gas and dark matter particles; for a subset of the
codes, we have full-physics (FP) runs, which include both stars, gas, and dark
matter particles, and a range of baryonic physics,
including gas heating and cooling, star formation, black hole growth, and
various sources of feedback.

Following on from the findings presented in \citetalias{Sembolini2015}, we separate
NR runs into two groups -- those run with codes that recover declining entropy
profiles with decreasing radius (classic SPH), which we refer to as ``Classic
SPH'', and those run with codes that recover entropy cores at small
radii (mesh, moving mesh, and modern SPH), which we refer to as ``non-Classic
SPH''. Further, we separate FP runs into runs with and
without black hole growth and AGN feedback (AGN and noAGN respectively).
The AGN feedback is believed to be essential for galaxy clusters, which
can solve the over-cooling problem, and provide better agreements with
observational results \citep[e.g.][and references
 therein]{Puchwein2008,Fabjan2010,Planelles2014,Planelles2015}. Although
\gadgetpesph\ does not directly include the AGN feedback, it uses the heuristic
model \citep{Rafieferantsoa2015} to quench star formation in massive
galaxies, which can be viewed as mimicking AGN feedback. Thus, we include
\gadgetpesph\ in the AGN instead of noAGN subgroup. For reference, we list all
simulation codes and implemented baryonic physics models in Table
\ref{tab:codes}. We summarise the key features of the codes that are relevant
for this study in appendix~\ref{A:codes}. We refer the reader to nIFTy
\citetalias{Sembolini2015,Sembolini2015b,Elahi2015} for a more detailed summary.

%%%%%%%%%%%%%%%%%%%%%%%%%%%%%%%%%%%%%%%%%%%%%%%%%%%
\section{The Simulated Galaxy Cluster} \label{sec:simulated_cluster}
%%%%%%%%%%%%%%%%%%%%%%%%%%%%%%%%%%%%%%%%%%%%%%%%%%%

We use the same massive galaxy cluster simulated in \citetalias{Sembolini2015,Sembolini2015b,Elahi2015} with a
virial mass of $M_{200}\simeq 1.1 \times 10^{15} h^{-1} \rm{M}_{\sun}$ and
virial radius of $R_{200} \simeq 1.69 \mpc$ at $z$=0 \footnote{$R_{200}$ is
the radius within which the enclosed mean matter overdensity is 200 times the
critical density of the Universe, while $M_{200}$ is the total mass within
$R_{200}$.}. This was selected from the \emph{MUSIC-2} sample
\citep{Sembolini2013, Sembolini2014,Biffi2014}, a dataset of hydrodynamical
simulations of galaxy clusters that were re-simulated from the parent
\emph{MultiDark}\footnote{www.cosmosim.org} dark-matter-only cosmological
$N$-body simulation \citep{Prada2012}. In these simulations, cosmological
parameters of $\Omega_{\rm M}=0.27$, $\Omega_{\rm b} = 0.0469$,
$\Omega_{\Lambda}= 0.73$, $\sigma_8=0.82$, $n = 0.95$, and $h = 0.7$
were assumed, in accordance with the WMAP7+BAO+SNI dataset presented in
\citet{Komatsu2011}.

The initial conditions of all the clusters of the \emph{MUSIC-2} dataset are publicly
available\footnote{\small{CLUSTER\_00019} of the \emph{MUSIC-2} sample at
  \texttt{http://music.ft.uam.es}}. Briefly, these were produced using
the zooming technique described in \cite{Klypin2001}. All particles within a
sphere with a radius of $6 \mpc$ around the centre of the halo in
the parent \emph{MultiDark} simulation at $z$=0 were found in a low-resolution
version ($256^3$ particles) of the parent, and mapped back to the parent's
initial conditions to identify the Lagrangian region from which these particles
originated. The initial conditions of the original simulations
were generated on a finer mesh of size $4096^3$. By doing so, the mass
resolution of the re-simulated objects was improved by a factor of 8
with respect to the original simulations. In the high resolution
region the mass resolution for the dark matter only simulations corresponds to
m$_{\rm DM} = 1.09 \times 10^9 \hMsun$, while for the runs including a baryonic
component, m$_{\rm DM} = 9.01 \times 10^8 \hMsun$ and m$_{\rm gas} =
1.9 \times 10^8 \hMsun$. In this paper, all the codes used the same aligned
parameters \citepalias[see the Table 4 in][]{Sembolini2015} to re-simulate the
selected cluster.

In our analysis, the cluster is first identified with AMIGA's-Halo-Finder
\citep[AHF]{AHF1,AHF2} and then its centre is defined as the position of the
minimum of the gravitational potential \citep[see][for discussion about the
agreement between different centre definitions]{Cui2015}. All the cluster
properties, such as, spherical overdensity mass, radial profiles, are
recalculated with respect to the minimum of the potential.

%%%%%%%%%%%%%%%%%%%%%%%%%%%%%%%%%%%%%%%%%%%%%%%%%%%
\section{Radial Profiles} \label{sec:pro}
%%%%%%%%%%%%%%%%%%%%%%%%%%%%%%%%%%%%%%%%%%%%%%%%%%%

\subsection{Mass Profiles}
\label{sec:DP}

%%%%%%%%%% FIG 1 & 2 %%%%%%%%e
\begin{figure*}
\includegraphics[width=1.\textwidth]{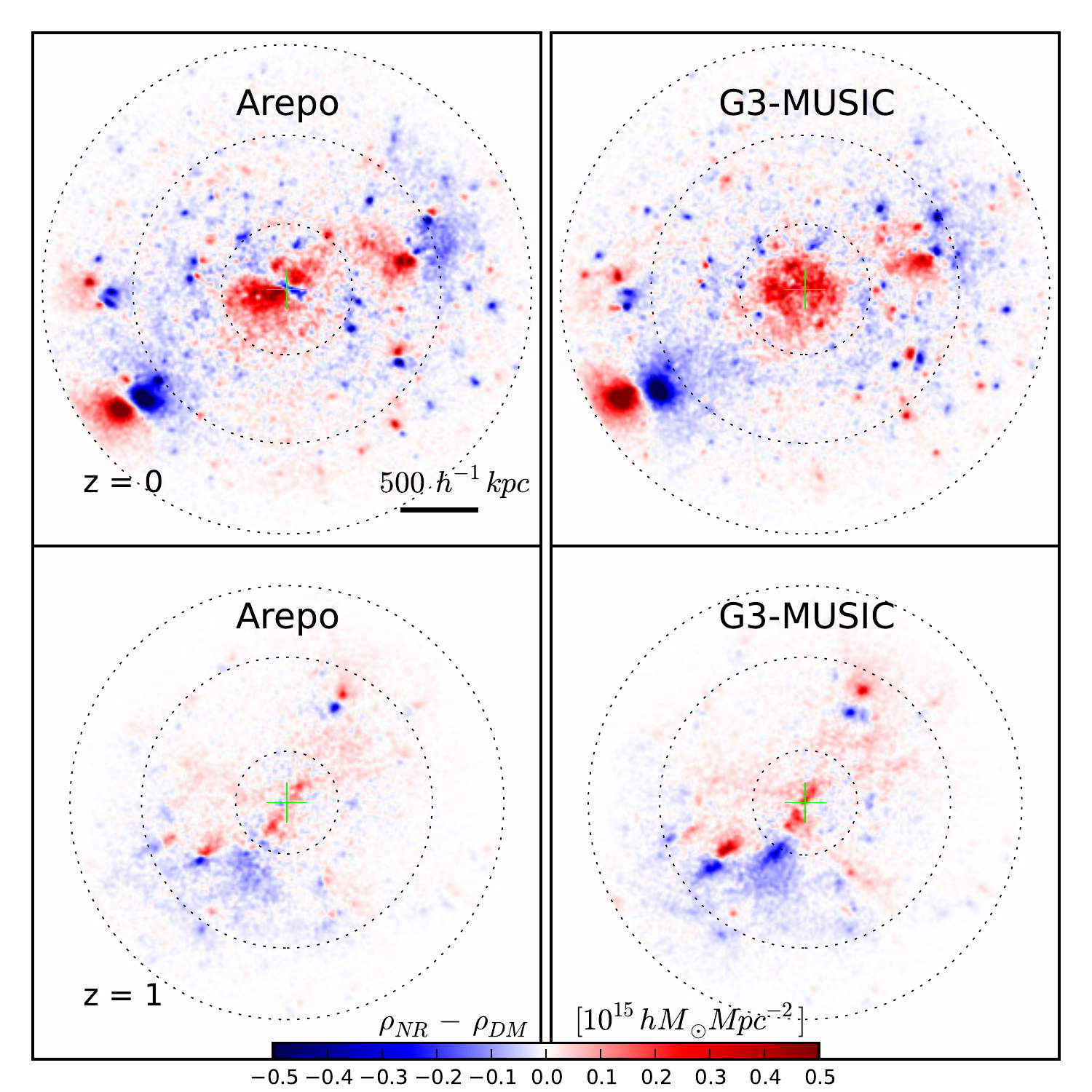}
\caption{Projected dark matter density difference between DM and NR runs. We
  only show two simulation codes -- \arepo\ and \gadgetmusic\ for illustration
  here. The colour is coding for the projected density difference, from negative
  values (blue) to positive values (red). The white region indicates no difference
  between the two runs. The simulation code name is shown on the top centre. The
  lime green cross in each plot indicates the aligned cluster centre position.
  The results in the upper (lower) row are from redshift $z$=0 ($z$=1). From
  inner to outer region, the three dotted circles represent $R_{2500}, R_{500}$
  and $R_{200}$ in the DM runs, respectively.}
\label{fig:show_nr}
\end{figure*}

\begin{figure*}
\includegraphics[width=1.\textwidth]{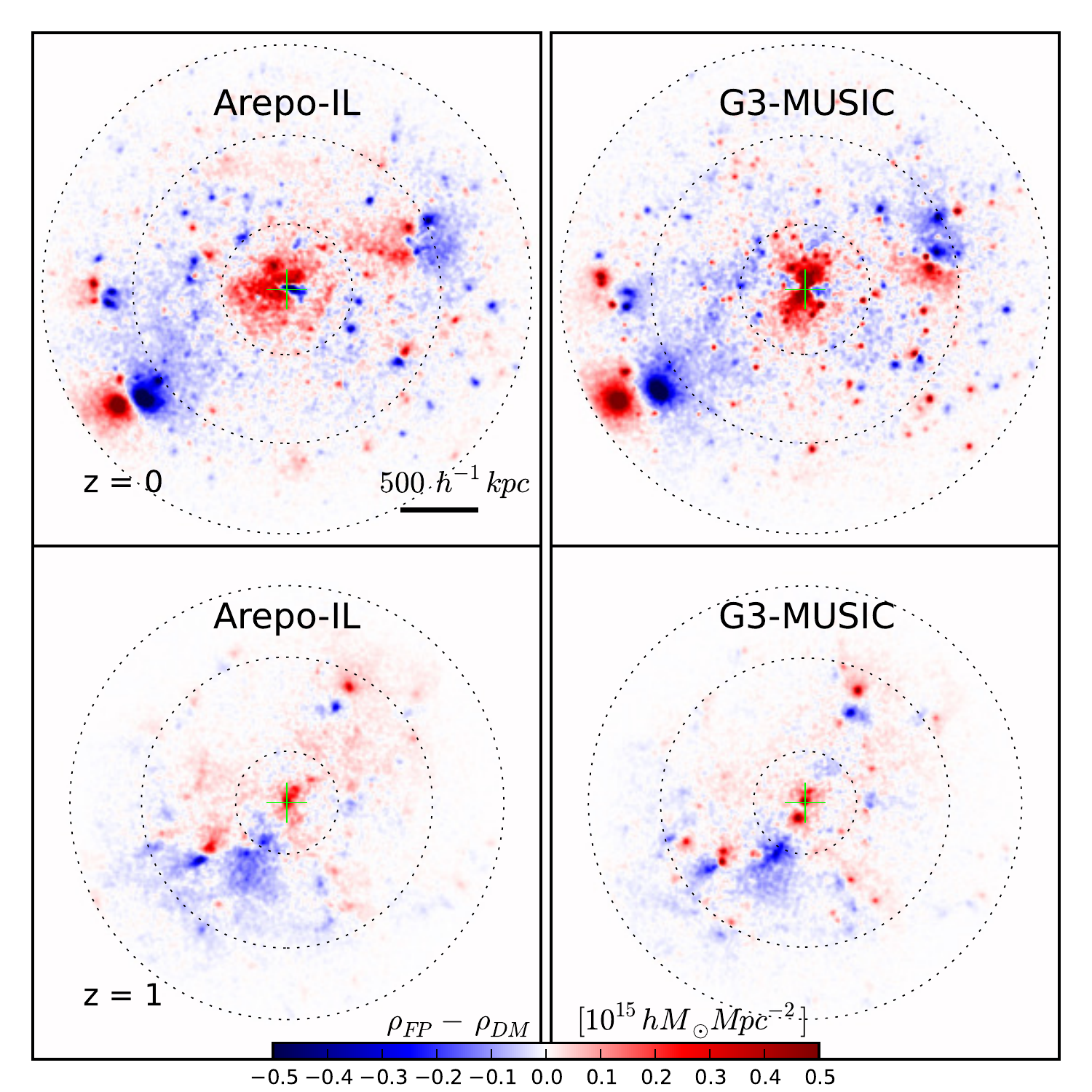}
\caption{Projected dark matter density difference between DM and FP runs.
  Similarly to Fig. \ref{fig:show_nr}, we only show two sample simulation
  codes -- \arepoil\ and \gadgetmusic\ here. Refer to Fig. \ref{fig:show_nr}
  for the details.}
\label{fig:show_fp}
\end{figure*}
%%%%%%%%%%%%%%%%%%%%%%%%%

\paragraph*{Visual Impression:} We begin by inspecting the differences in
projected dark matter density between the DM and NR runs shown in
Fig. \ref{fig:show_nr}, and between the DM and FP runs, shown in
Fig. \ref{fig:show_fp}. Here we show two examples from simulation codes
drawn from the ``Classic SPH'' and ``non-Classic SPH'' subgroups --
respectively, \gadgetmusic\ and \arepo. In practice, we use only the
high-resolution dark matter particles within $R_{200}$ and compute densities
using a standard cubic spline SPH kernel with 128 neighbours at the position of
each dark matter particle; these densities are then smoothed to a 2D mesh (on
x-y plane with a pixel size of $5 \kpc$)
using the same SPH kernel \citep{Cui2014b, Cui2015}. To show the
projected dark matter density difference, these images are simply aligned
with the cluster centre without further adjustment. The density change is
given by $\delta_{\rho} =
\rho_{NR,FP} - \rho_{DM}$; in Fig.~\ref{fig:show_nr}, blue (red) indicates a
negative (positive) $\delta_{\rho}$, or depressed (enhanced) densities in the
NR and FP runs with respect to the DM run. Note that dark matter particles have
a slightly larger mass in DM runs than in the NR and FP runs; we compensate for
this by correcting the dark matter particle mass in NR and FP runs to be
the same as in the DM run.

Fig.~\ref{fig:show_nr} clearly shows that, at $z$=0, dark matter density
changes are normally within $0.5 \times 10^{15}~h~{\rm M_{\sun}~Mpc^{-2}}$ over all
the cluster, except within the central
regions and at the positions of satellites. In the centre, the dark matter
density is depressed relative to the DM runs in the non-classic SPH runs,
as shown in the \arepo\ panels, while the majority of classic SPH codes
showed enhanced central densities, as shown in the \gadgetmusic\ panels.
The density variations associated with substructures are also evident,
especially at $z$=0 associated with the large infalling substructure (to the
bottom-left) on the outskirts of the cluster, indicating that the inclusion
of gas can introduce an offset in the timing of mergers between DM and NR
runs. At redshift $z$=1, differences in density are smaller than at $z$=0,
and the enhanced density within the central regions is evident in both subgroups
of codes.

In Fig.~\ref{fig:show_fp}, we show how the dark matter density changes
between the DM and FP runs, and see similar trends as in Fig.
\ref{fig:show_nr}. Interestingly, the additional baryonic processes, most
likely gas cooling, in the FP runs compared to the NR runs result in
obvious density contrasts within the central regions and in substructures.
It's important to note at this point, and we shall make this clear in the
remainder of the paper, that the split into the classic and non-classic SPH
groupings is not really appropriate for the FP runs; there are large code
to code variations within these subgroups, primarily driven by the baryon physics implementations.

%%%%%%%%%% FIG 3 %%%%%%%%
\begin{figure*}
\includegraphics[width=1.0\textwidth]{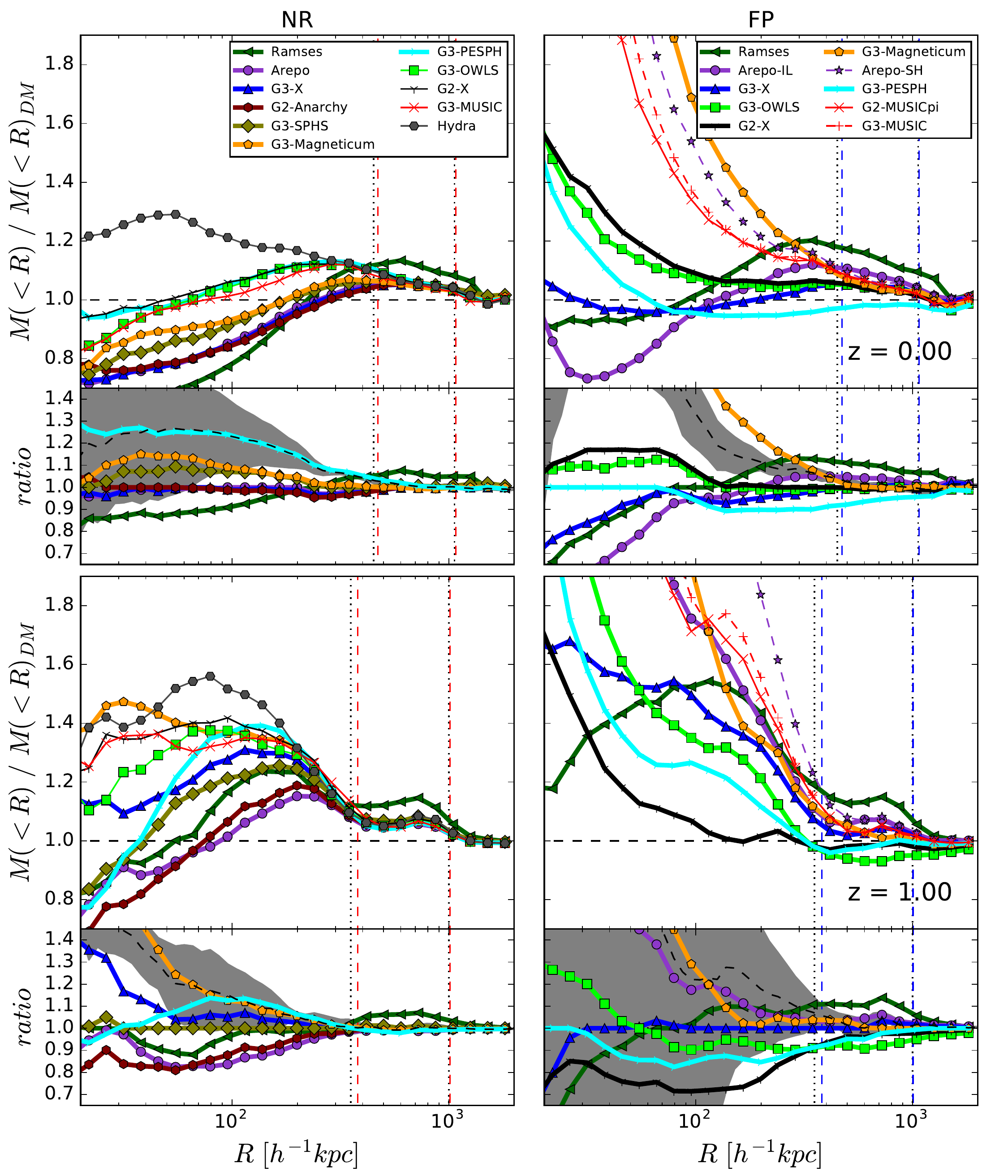}
\caption{Differences in the cumulative mass profile between the NR/FP and DM
  runs. The left column shows the difference between the mass profile in the
  NR and DM runs, while the right column shows the corresponding result for the
  FP and DM runs. The line
  style, colour and symbol for each code are indicated in the legend.
  Vertical dashed (red for the NR runs; blue for the FP
  runs) lines show $R_{2500}$ (inner) and $R_{500}$ (outer) from the
  \gadgetmusic\ runs, while vertical dotted black lines are from the DM run.
  We show the results at $z$=0 (top panel) and $z$=1 (bottom panel). Under
  each plot, we show the residuals with respect to the median of the
  non-classic SPH density
  profiles (or the median profile of the AGN subgroup in the right column),
  which are also shown in thick lines in the upper panels. The thin black dashed
  lines are the median profiles from classic SPH codes (or the median profiles
  from the noAGN subgroup in the right column) with 1-$\sigma$ error
  shown by the shadow region. The classic SPH codes (also the noAGN codes in the
  right column) in the upper panel are shown in thin lines.}
\label{fig:adp}
\end{figure*}
%%%%%%%%%%%%%%%%%%%%%%%%%

\paragraph*{Total Enclosed Mass Profiles}
In Fig.~\ref{fig:adp} we show how the enclosed total mass density profile varies
between the NR and DM runs (left column) and FP and DM runs (right column)
at $z$=0 (upper panels) and $z$=1 (lower panels). We use fixed size in
logarithm for each radial bin. Within each panel, we
show the radial profiles (upper section) and the residuals with respect
to the median profiles (lower section). Vertical lines denote $R_{2500}$ and
$R_{500}$ measured in the fiducial \gadgetmusic\ DM (black dotted lines) and
corresponding NR and FP runs (red and blue dashed lines respectively). A lower
cut of $R = 20 \kpc$, roughly in accordance with the convergence criterion
presented in \cite{Power2003} has been applied. The data are separated according
to the classic and non-classic SPH classification (thin and thick curves) in
the case of the NR runs, and the AGN and noAGN classification (thick and
thin curves) in the case of the FP runs.
%  We note here that there is no \ramses\
% data at $z$=1 in Fig.~\ref{fig:adp} and all following figures.

We have already seen evidence in Fig.~\ref{fig:show_nr} that the dark matter
density in the central regions of the cluster is depressed in the NR runs
relative to the DM runs at $z$=0. This
depression is evident in the total spherically averaged profiles;
the non-classic SPH codes show densities of $\sim 80$ \% of
their value in the DM run in the central regions of the cluster, while the
classic SPH codes show a greater variation, ranging from a density of
$\sim 80$ \% of the DM value for \gadgetmusic\ to $\sim 120$ \%
for \hydra. Similar behaviour as the classic SPH code -- \gadgetmusic, has been
reported in \cite{Cui2012a}; see Fig. 4 in this paper for more details.

The density
is enhanced in the NR runs relative to the DM runs at large radii, outside of
$R_{2500}$, in all of the codes. Interestingly, at $z$=1, this
trend of an enhancement in density continues to small radii, before plateauing
and in some cases inverting, so that the density is
depressed in the NR run relative to the DM run; notably, the codes that
invert and show density depressions relative to the DM run are all non-classic
SPH codes. At $z$=1, it is also
noticeable that the variation between codes is large at small radii; the change
is $\sim 20 - 50$ \% at 100 $\kpc$. At $z$=0, the variation is much
smaller, $\sim 20$ \% at 100 $\kpc$, $\sim 30$ \% if we include the
outlier, \hydra. The mesh code \ramses\ shows larger increases respected to its
DM run between $R_{2500}$ and $R_{500}$ than all the other codes at both $z$=0 and 1.
It means that this difference can be traced back to even high redshift.
The non-classic SPH code \gadgetpesph\ has the largest
deviation with respect to other non-classic SPH codes. It shows a similar
behavior as the classic SPH code \gadgettwox, which could be caused by a
convergence issue \citep{Read2012}.

To highlight the scatter between different codes, we show residuals with
respect to the median for each of the non-classic SPH codes as individual
curves in the lower panels, while we show residuals with respect to the median
for the grouped classic SPH codes as the median (black dashed curves) and
1-$\sigma$ variation (shaded region). This shaded region is only indicating the
scatter between the classic SPH codes. For example, its lower boundary
does not mean that the classic SPH codes have the possibility of producing
such low density. The disparity between the
median values of the classic and non-classic SPH codes can be seen at
$R \simlt R_{2500}$ at both redshifts. The difference between the two subgroups is
as large, if not larger than, the scatter between the codes within each
subgroup; classic SPH codes tend to have roughly 20 \% higher central
densities than non-classic SPH codes. It is worth to note here that the
agreement between non-classic SPH codes at $z$=0, can not be reached at $z$=1,
which shows a larger scatter $\sim$ 50 \%.

\medskip

The impact of baryonic physics on the total mass profile is particularly
striking in the FP runs, with large variations between the different codes.
At $z$=0, the density within $R_{2500}$ is enhanced in the majority of the
codes, with only \ramses, \arepoil\ and \gadgetx\ showing depressed
densities. It is
interesting that all three noAGN runs show increasing enhancements in
relative density with decreasing radii, whereas there is no clear trend
in the AGN runs, with some showing depressed relative central densities
while others show strong enhancements. At $100 \kpc$, the densities in the AGN
runs relative to the DM runs vary between $\sim 100$ \% to $\sim 180$ \%, while
the noAGN runs have relative densities varying between $\sim 130$
\% to $\sim 160$ \%.
%depending on whether or not \gadgetpesph\ is included.
At $z$=1, all of the runs show relative
density enhancements within $R_{2500}$, ranging from $\sim 100$ \% to
an excess of $200$ \%; as at $z$=0, %if we ignore \gadgetpesph\,
then we see that the three other noAGN runs show the largest relative
enhancements at all radii.
At $z$=0, \gadgetmagneticum\ produces the largest enhancement within $R_{2500}$ in its FP run, however mimicking the behaviour of the other AGN codes in outer region and at redshift $z$=1. This could be caused by the specific implementation of AGN feedback model, where BH merging and the parameters regulating the accretion onto the BH and the associated feedback are treated differently \citep[see more details in][]{Steinborn2015}.
Although \gadgetpesph\ does not directly include the
AGN feedback, it shows a similar behaviour as the AGN codes \gadgettwox\ and
\gadgetowls\/ \citepalias[see also in][]{Sembolini2015}. This could
be caused by its highly constrained heuristic model for galactic outflows
\citep{Dave2013}, which utilises outflows that scale as momentum-driven
winds in sizeable galaxies.

The large variations in the behaviour of the curves in the AGN and noAGN runs
with respect to the median, as shown in the residuals, emphasises the trends
we have just noted. At $R \simgt R_{200}$, there is a good agreement between
all of the codes for both AGN and noAGN runs; for $R_{200} \simgt R \simgt R_{2500}$, the
differences become pronounced -- up to $\sim 0-20$ \% -- again regardless of
whether or not they are AGN or noAGN. It is worth to note that the \ramses\ still
has the highest enhancement compared to the other codes as its NR run; while at $R \simlt R_{2500}$, the
variation with respect to the median is striking, especially in the case of the
AGN runs. This is true at both $z$=0 and $z$=1.

\medskip %%discussion

These trends are consistent with the results of \citet{Martizzi2012}, with mass
profiles from the FP runs close to DM runs ($\simlt 20$ \%) at radii
$R \simgt 0.1 \times R_{200}$, and with \cite{Lin2006} and \citet{Cui2012a},
who also found lower relative central densities in the NR runs. The non-classic
SPH codes tend to
have lower central relative densities when compared to the classic SPH
counterparts; because of gas pressure and energy redistribution between
dark matter and gas particles during halo collapse, all the codes show a
relative density enhancement at $R_{500} \simgt R \simgt 200 \kpc$ (this value is much
smaller for the classic SPH codes and for the higher redshift). Similar results
have been found in \cite{Rasia2004}, \citet{Lin2006}, and \citet{Cui2012a}.

The sensitivity of relative central densities to baryonic physics -- of the
kind implemented in the FP runs -- has been reported previously \citep[e.g.][]{Duffy2010,Teyssier2011,Martizzi2012,Cui2012a,Cui2014a,Velliscig2014,Schaller2015a}. What is
particularly interesting about our results is how much variation is evident
in runs that seek to implement broadly similar baryonic physics prescriptions,
especially at $z$=0. Such variation is consistent with previous work; some
studies report on enhancements in relative central densities,
consistent with the \gadgettwox, \gadgetmagneticum\ and \gadgetowls\ AGN runs
\citep[e.g.][]{Duffy2010, Cui2014a, Velliscig2014}, while others report
on relative central density depressions consistent with the \ramses,
\arepoil\ and \gadgetx\ AGN runs \citep[e.g.][]{Teyssier2011, Martizzi2012}.
Understanding this variation is not straightforward -- not only do the
precise baryonic physics implementations differ, but there are also
differences in the underlying scheme to solve the equations of gas dynamics,
as the split between classic SPH codes, such as \gadgettwox\ and \gadgetowls,
and non-classic SPH codes, such as \gadgetmagneticum\ and \gadgetx\ highlights.

\subsection{Kinematic Profiles}
\label{kinematic_profiles}

%%%%%%%%%% FIG 11 %%%%%%%%
\begin{figure*}
\includegraphics[width=1.0\textwidth]{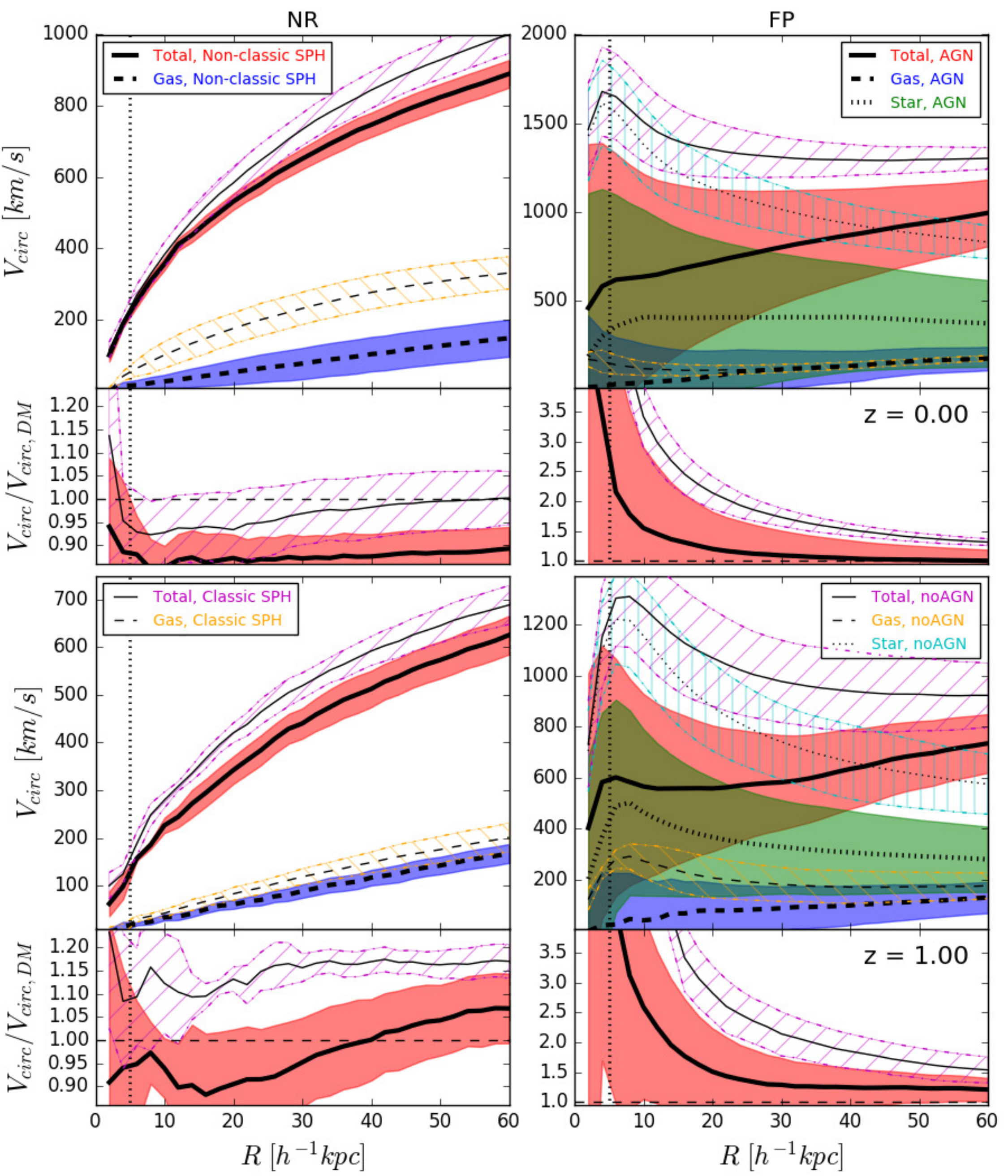}
\caption{The circular velocity profile at the centre of the simulated cluster
  from NR runs (left panel) and FP runs (right panel).
  As indicated in the legends, the solid lines show the total circular velocity
  in the cluster centre; the dashed lines are the circular velocity from the gas
  component; the dotted lines are from the stellar component; the different
  coloured regions / hatchings and lines of different width show the standard
  deviation and median profile between
  different simulation codes in each subgroup, as indicated in the legends. The
  lower subplot below each main panel shows the total circular velocity
  difference between the NR/FP and DM runs. From top to bottom, we show
  the results at $z$=0 and $z$=1. The vertical dotted lines show the softening
  length in the simulation.}
\label{fig:vcs}
\end{figure*}
%%%%%%%%%%%%%%%%%%%%%%%%%

%%%%%%%%%% FIG 12 %%%%%%%%
\begin{figure*}
\includegraphics[width=1.0\textwidth]{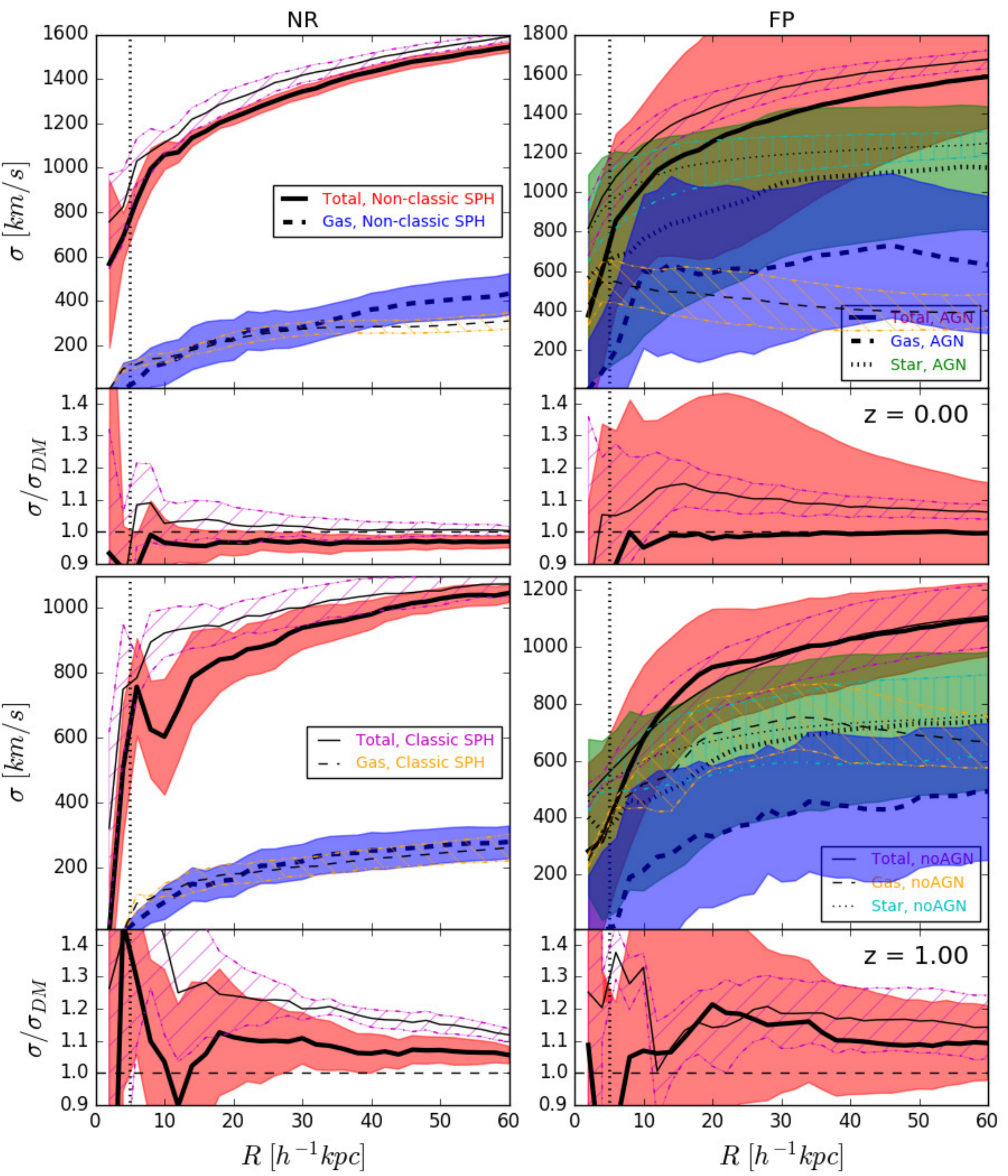}
\caption{Similarly to Fig. \ref{fig:vcs}, but for the velocity dispersion
  profile at the centre of the simulated cluster. Refer to Fig. \ref{fig:vcs}
  for more details of the subplot distributions and to the legends for the line
  styles and coloured region / hatching meanings.}
\label{fig:vds}
\end{figure*}
%%%%%%%%%%%%%%%%%%%%%%%%%

The previous results highlight that the inclusion of baryons has a
significant impact on the mass distribution within the simulated cluster,
especially within the central regions. We now investigate how this influences
kinematic profiles.

\paragraph*{Circular Velocity:}
In Fig.~\ref{fig:vcs}, we show how the circular velocity
profile within the cluster centre ($R \leq 60 \kpc$) varies between the
NR and DM runs (left column) and FP and DM runs (right column)
at $z$=0 (upper panels) and $z$=1 (lower panels). We limit the profile within
$60 \kpc$ because we are interesting in the core region of $R_{2500}$, where
the profile is dominated by the BCG in the FP sims. A fixed linear radial bin
size is applied here. Within each panel, in the upper section we show the median
profiles of the total matter (solid curves), gas (dashed curves), and, if present,
stars (dotted curves), with the shaded regions and the hatchings between dot-dashed
lines indicating the 1-$\sigma$ variation with respect to the median; in the lower
section we show residuals with respect to the corresponding total matter profiles in the DM runs.
Vertical lines denote a gravitational softening length of $5 \kpc$, which was
used in the DM and NR runs, and which is used as indicative of the softening
in the FP runs. NR runs are grouped into non-classic SPH (thick lines with
red shadow region) and classic SPH (thin lines with magenta shadow region),
while FP runs are grouped into AGN (thick lines with red shadow region) and
noAGN (thin lines with magenta shadow region) runs.

The residuals are particularly instructive. For the NR runs, at $z$=0, there is
a $\sim$ 1-5 \% change in the total matter circular velocity in the
classic SPH runs compared to DM runs, $\sim$ 10-15 \% lower for the
non-classic SPH runs; the change in circular velocity of the gas component
between the classic and non-classic SPH runs is significant, in excess of
100 \%. At $z$=1, the classic SPH total matter circular velocity profile
is $\sim$ 15 \% higher than in the DM runs, whereas the non-classic SPH
total circular velocity changes by between $\sim$ -10 to +10 \% from
the inner to outer radius; the circular velocity profiles of the gas
components are now much more in agreement with one another, differing
by $\sim 10$ \% at most.

In the case of the FP runs, the impact of baryonic physics on the total matter
circular velocity profile is substantial, with enhancements by factors of
$\sim 1.5 (3.5)$ at $10 \kpc$ and quickly decreasing to $\sim 0 (40)$ \% at $\sim 60 \kpc$, relative
to the circular velocity profiles in the DM runs at $z$=0 for the AGN (noAGN) subgroup. The enhancements are
greatest for the noAGN runs, as we might expect -- without the influence of the
AGN, gas cooling can proceed relatively unhindered. There are significant
differences between the stellar circular velocity profiles in the noAGN and AGN
runs at both $z$=0 and $z$=1, by a factor of $\sim 2-3$ over the radial range,
whereas the differences between the gas circular velocity profiles are
comparatively small -- there is good consistency between the AGN and noAGN runs
at $z$=0, although the noAGN profile is about tens of per cent higher than the
AGN profile at $z$=1.

\medskip

\paragraph*{Velocity dispersion profiles:}
As in Fig. \ref{fig:vcs}, we show the total matter (solid line), gas (dashed
line) and stellar (dotted line) velocity dispersion ($\sigma$) profiles from
both NR and FP runs (left and right columns respectively) in Fig.
\ref{fig:vds}. In the upper (lower) panels we show results from $z$=0 (1),
and in the upper (lower) section we show the differences with
respect to the velocity dispersion profile in the corresponding DM run. The data
is also binning in the same fixed linear size as in Fig. \ref{fig:vcs}.

In the case of the NR runs, the total velocity dispersion profiles in
the classic SPH and non-classic SPH runs are in very good agreement at $z$=0
and reasonable agreement at $z$=1. At $z$=0, the difference with respect
to the DM runs is small, with the ratio of $\sigma/\sigma_{DM}$ of order
unity; at $z$=1, the difference is slightly greater, showing an enhancement by
a factor of $\sim 1.1-1.3$ greater than in the DM run (greater within
$\sim 10 \kpc$). The gas velocity dispersion profiles are broadly similar
in the classic and non-classic SPH runs at both redshifts.

Against the circular velocity profiles in the FP runs, here we see
a less significant variation in the velocity dispersion profiles with respect to
the median, evident in Fig. \ref{fig:vds}. At $z$=0, the median total matter
and stellar velocity dispersions have a broadly similar shape and amplitude,
albeit with the noAGN velocity dispersions being larger; the gas profiles
show a slightly larger discrepancy, although both are flat over most of the radial
range, and here the AGN velocity dispersion is larger, as we might expect
in the presence of feedback from the central AGN. Relative to the DM runs
velocity dispersion profiles, we see that the ratio with respect to both
AGN and noAGN is flat and of order unity in the AGN runs and $\sim 1.1$
in the noAGN runs. Both sets of runs show a decline within the central
$\sim 10 \kpc$. At $z$=1, the total matter velocity dispersion in the AGN
runs rises sharply in the inner regions before flattening off at $R \simgt 30
\kpc$, whereas the noAGN case shows a steady increase with increasing
radius. The difference with respect to the DM run is shown in the lower
section, and we see that the ratio in both the AGN and noAGN runs is flat
at $R \simgt 20 \kpc$ and corresponds to an enhancement by a factor of
$\sim 1.2$, but shows a smaller enhancement in the AGN run and a slightly larger
enhancement in the noAGN run at $R \simlt 20 \kpc$. The gas velocity dispersion
profiles show an inversion of the behaviour evident at $z$=0 with large difference;
while the stellar velocity dispersion differences between the median values from
the AGN and noAGN groups are smaller compared to the $z$=0 result.

The circular velocities for the gas, stellar and total components from the AGN
subgroup are similar to the results from \cite{Schaller2015a} (see the most massive
groups in the Fig. 6 for details).
By comparing their non-radiative simulation with the one including gas cooling and stellar feedback,
\cite{Lau2010} showed that the baryon dissipation increases the velocity dispersion of
dark matter within the virial radius by $\approx 5 - 10\%$. This effect is mainly
driven by the changes of the density and gravitational potential in inner regions of cluster.
Their explanation for the changes in the velocity dispersion is explicitly shown in Fig. \ref{fig:vds}.

%%%%%%%%%%%%%%%%%%%%%%%%%%%%%%%%%%%%%%%%%%%%%%%%%%%
\section{Global Properties} \label{sec:gp}
%%%%%%%%%%%%%%%%%%%%%%%%%%%%%%%%%%%%%%%%%%%%%%%%%%%

\subsection{Enclosed Mass}

%%%%%%%%%% FIG 4 %%%%%%%%
\begin{figure*}
\includegraphics[width=1.0\textwidth]{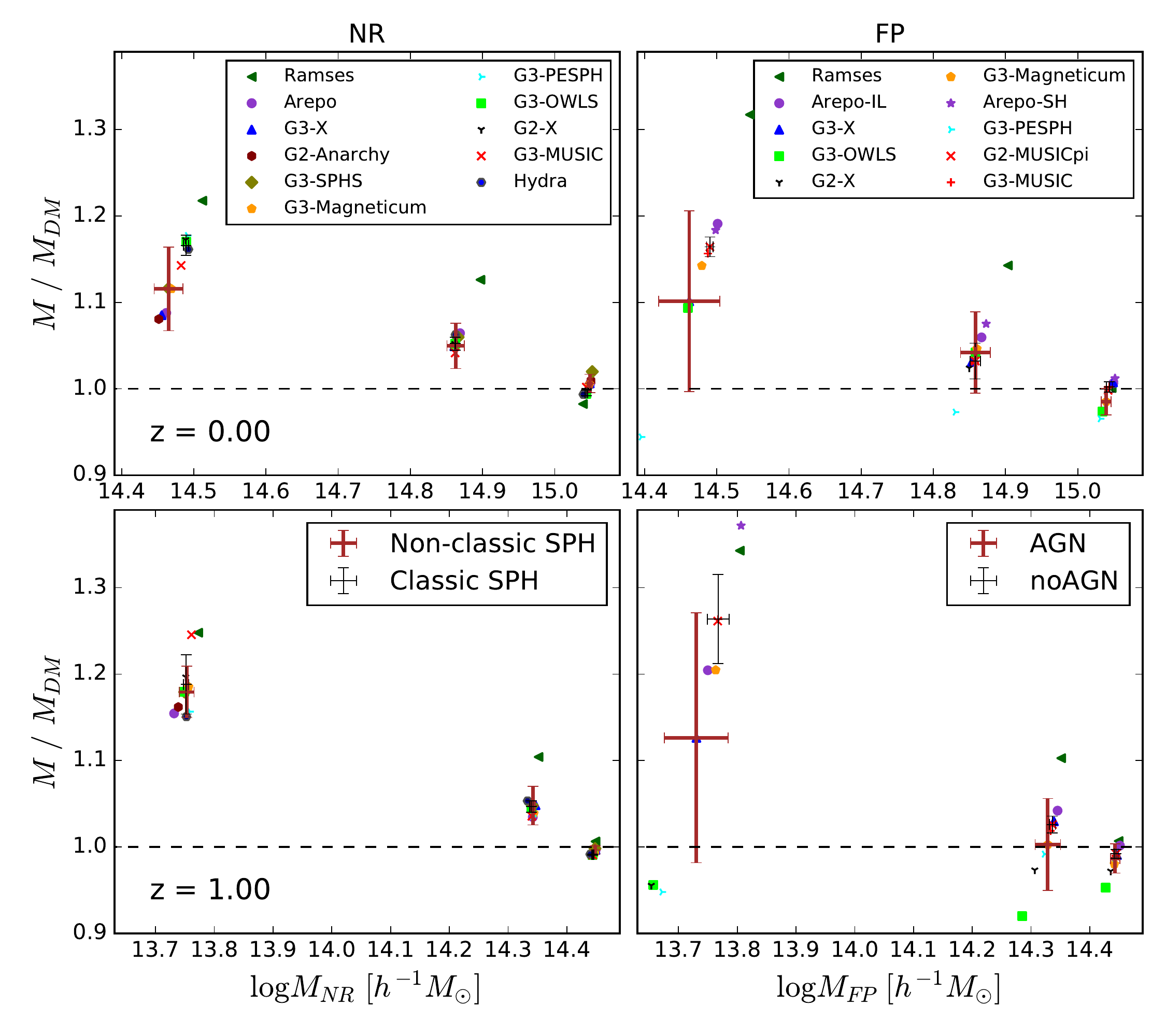}
\caption{Halo mass difference between the DM runs and the NR runs (left column)
  / the FP runs (right column). As indicated in the legends in the top row,
  different coloured symbols indicate different simulation codes. In each panel,
  there are three groups of data with error bars, which correspond to
  $M_{2500}, M_{500}$ and $M_{200}$ from smaller to larger halo mass. The
  meaning of the error bars in both columns is shown in the legends in the two
  lower panels: the brown thick one is for the non-classic SPH subgroup (AGN
  subgroup in the right column); while the black thin errorbar is for the
  classic SPH subgroup (noAGN subgroup in the right column). From top to bottom
  panel, we show the results at $z$=0 and 1, respectively.}
\label{fig:md}
\end{figure*}
%%%%%%%%%%%%%%%%%%%%%%%%%

As we saw in Fig.~\ref{fig:adp}, there are mass profile changes at
$R_{2500}, R_{500}$ and $R_{200}$. These changes are directly connected to the
spherical overdensity (SO) halo mass. In Fig. \ref{fig:md}, we show how the
measured SO masses -- from left to right,
$M_{2500}, M_{500}$ and $M_{200}$ -- vary with respect to the
DM run in the NR runs (left column) and FP runs (right column) at
$z$=0 (upper panels) and $z$=1 (lower panels). The meaning of the different
coloured symbols is indicated in the insets.

The change in $M_{200}$ is negligible; $M_{NR, FP} / M_{DM} \approx 1$ with
whiskers indicating variations of $\pm 2$ \% at both redshifts,
independent of code used or baryonic physics implemented. The change in
$M_{500, NR}$ is already slightly larger, $\sim 5$ \% compared to $M_{500, DM}$,
at both redshifts; there is good consistency between codes in the classic
SPH and non-classic SPH, and AGN and noAGN subgroups, although the scatter
is larger in the FP runs. At the highest overdensity, $M_{2500}$, we see
the greatest mass increase with very large errorbars for both the NR and FP
runs and median enhancements of $\sim 10-20$ \%. In the NR runs,
there is a clear separation in the medians at both $z$=0 and 1 between the
classic and non-classic SPH runs, with the larger change in the classic SPH
runs, as the results so far imply; the variation with respect to the median
is smaller in the classic SPH runs, but it never exceeds $\sim 10$ \%.
In the FP runs, there is a large variation with respect to the median
in both the AGN and noAGN runs at both $z$=0 and 1, in excess of $\sim 10 (20)$
\% at $z$=0 (1); again, the trend is as we would expect, with the noAGN
runs having larger values of $M_{2500}$, arising from enhanced gas cooling and
star formation in the core.

\smallskip

The influence of baryonic physics on mass has been investigated by
a number of authors \cite[e.g.][]{Gnedin2004,Stanek2009,Cui2012a,Martizzi2013,Cusworth2013,Velliscig2014,Cui2014a,Bocquet2015,Khandai2015,Schaller2015a,Sawala2013,Chan2015,Zhu2015}. Our results are consistent with the findings of \cite{Cui2012a} (see
Fig. 2 for more details), and in particular, the insensitivity of $M_{200}$
to simulation code and precise baryonic model is in broad agreement with
previous studies \citep[cf. the work of ][who focused on cluster mass scales]{Cui2014a,Schaller2015a}.

\subsection{Central Density Profile}

%%%%%%%%%% FIG 10 %%%%%%%%
\begin{figure*}
\includegraphics[width=1.0\textwidth]{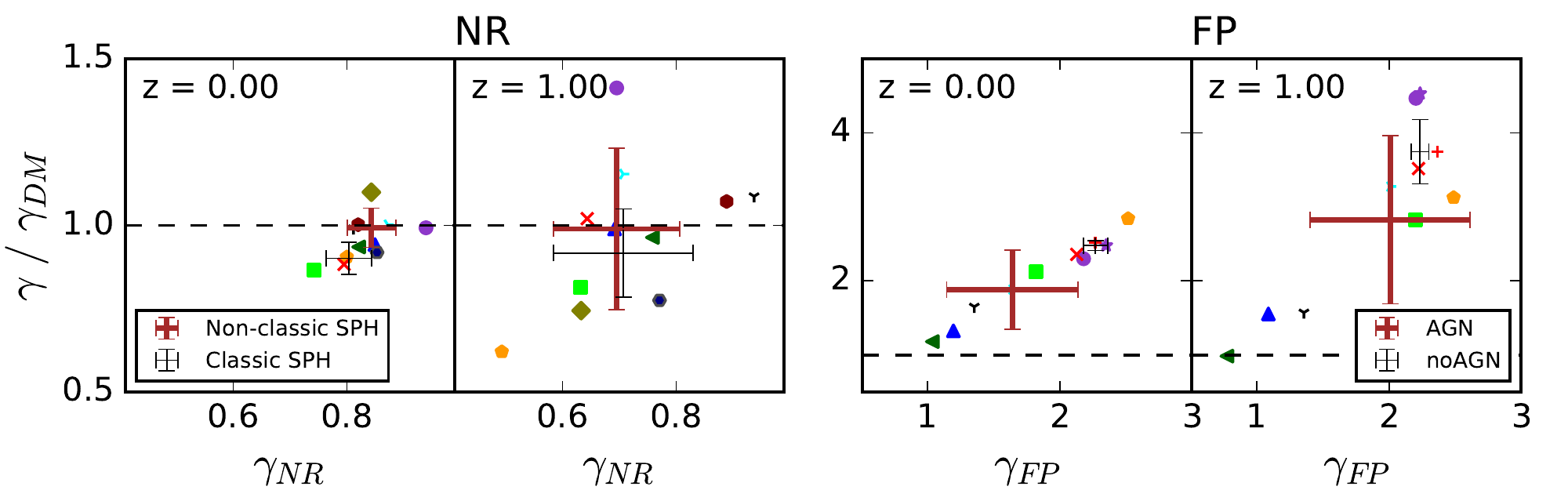}
\caption{The inner slope changes for the NR runs (left panel) and the FP runs
  (right panel). The coloured symbols represent the different simulation codes
  as in Fig. \ref{fig:md}. The meaning of the errorbars are shown in the legends
  in both left and right panels.}
\label{fig:inp}
\end{figure*}
%%%%%%%%%%%%%%%%%%%%%%%%%

Following \cite{Newman2013,Schaller2015b}, we characterise the central total
mass density profile by the average
logarithmic slope over the radial range $0.003 R_{200}$ to $0.03 R_{200}$,
\begin{equation}
\gamma = - <\frac{d \log \rho_{tot}(r)}{d \log r}>,
\end{equation}
here we used 25 equally spaced logarithmic bins to construct the density
profile. We have verified that the number of bins has little effect on the $\gamma$
value as long as it is larger than 10. The results are shown in Fig.
\ref{fig:inp} and reveal some interesting trends.

\smallskip

Firstly, the average slope in the NR runs increases from
$\gamma_{NR} \simeq 0.7$ at $z$=1 to $\gamma_{NR} \simeq 0.8$ at $z$=0,
while the variation in $\gamma_{NR}$ with respect to the mean decreases
by a factor of a few between $z$=1 and $z$=0.

Secondly, the ratio of the average
slope in the NR runs with respect to the DM runs shows little variation with
redshift -- $\langle \gamma_{NR}/\gamma_{DM} \rangle \simeq 1$ for the
non-classic SPH runs, $\langle \gamma_{NR}/\gamma_{DM} \rangle \simeq 0.9$
for the classic SPH runs -- whereas the variation with respect to the mean
shows a sharp decrease between $z$=1 and $z$=0, by a factor of several.

Thirdly, there is a large spread in slopes in the FP runs, ranging from
$\gamma_{FP}\simeq 1$ to 3, at both $z$=0 and $z$=1; separating
runs into those with and without AGN and taking the average reveals no
difference at $z$=1 ($\langle \gamma_{FP} \rangle \simeq 2.2$ for both AGN
and noAGN runs), whereas there is a reasonably significant difference at
$z$=0 ($\langle \gamma_{FP} \rangle \simeq 1.5$ for AGN runs,
$\langle \gamma_{FP} \rangle \simeq 2.2$ for noAGN runs) and in the sense we
might expect (i.e. steeper slopes in the noAGN runs, indicating enhanced
star formation and cold gas in the central galaxy). The median value from the
AGN runs is slightly higher than the result from \cite{Schaller2015b}.

Fourthly, there are
dramatic enhancements in the average slope in the FP runs with respect to
the DM runs, with $\langle \gamma_{FP}/\gamma_{DM} \rangle \simeq 2$ at $z$=0
and $\langle \gamma_{FP}/\gamma_{DM} \rangle \simeq 3-4$ at $z$=1, and as in
the NR runs, the variation with respect to these averages shrinks by a factor of
$\sim 2-3$ between the AGN and noAGN runs at both redshifts.

\subsection{Concentration}
%%%%%%%%%% FIG  %%%%%%%%
\begin{figure*}
\includegraphics[width=1.\textwidth]{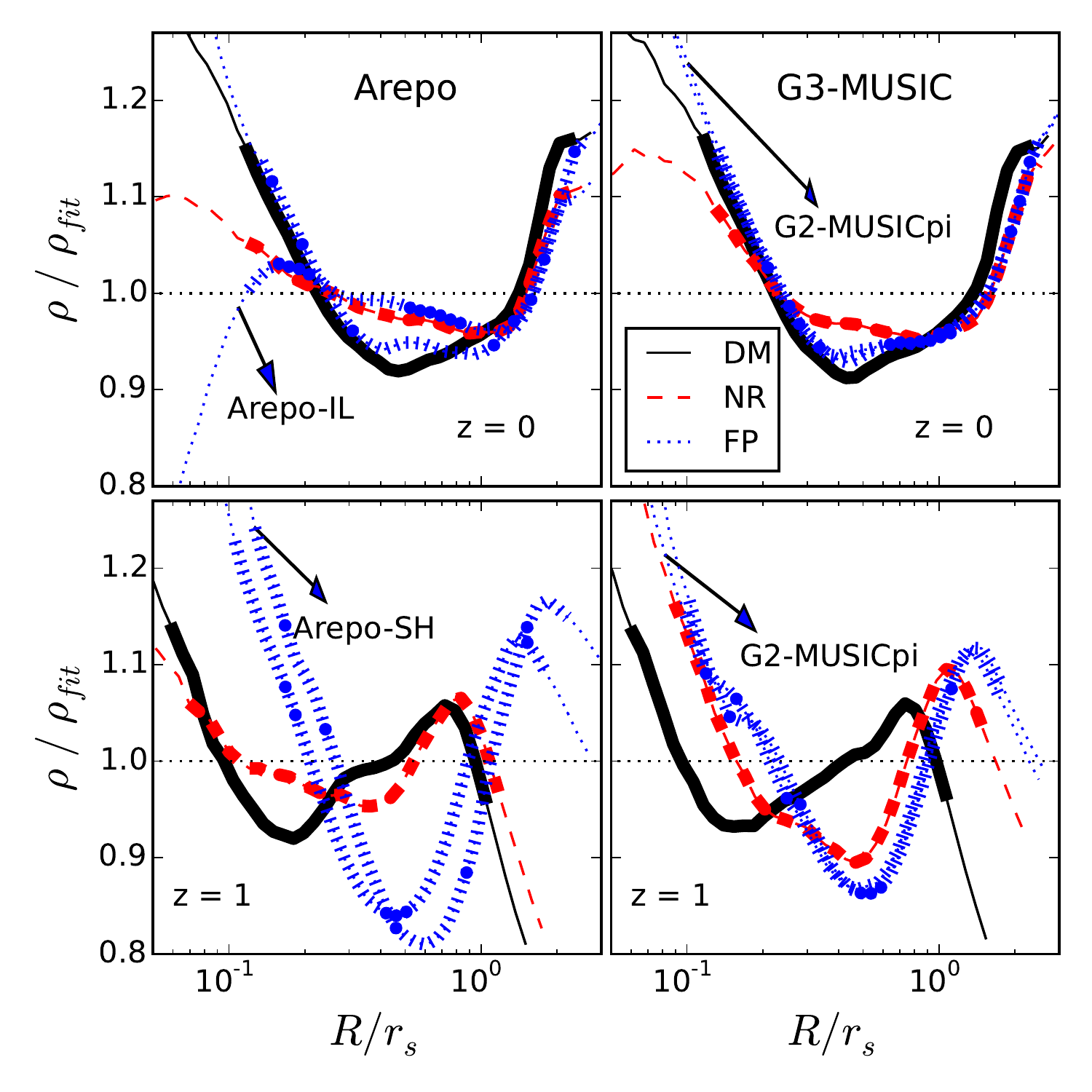}
\caption{Mass profile ratio to the NFW fitting for the dark matter component as a
  function of radius, which is normalized to the fitted parameter $r_s$. Similarly to
  Fig. \ref{fig:show_nr}, we only illustrate two example simulations at
  here. The simulation code names are shown in the top of each panel.
  Different color and style lines represent different baryonic models as
  indicated in the legend of the top-right panel. The thick lines indicate the
  region used for the NFW fitting. Upper row shows the result at $z$=0, while the
  lower row is the result at $z$=1.}
\label{fig:dmfit}
\end{figure*}

\begin{figure*}
\includegraphics[width=1.0\textwidth]{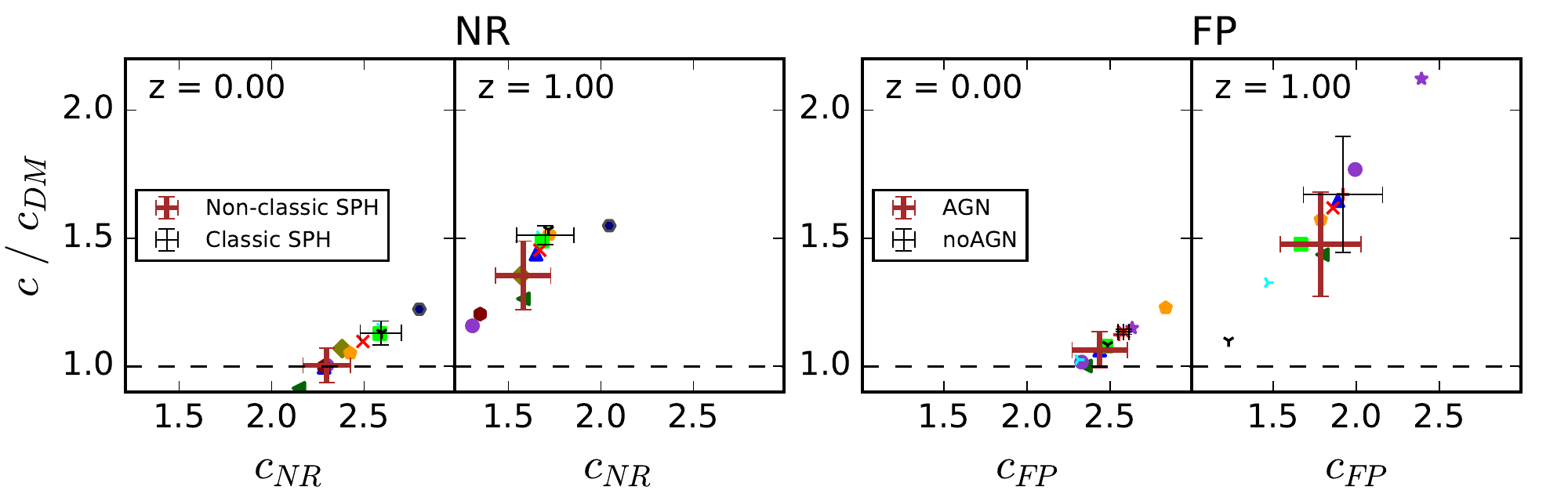}
\caption{Concentration changes with respect to the DM runs. The left column
  shows the results from NR runs, while the right column is from the FP runs.
  The two subplots in each column show the results at both $z$=0 and $z$=1,
  which is indicated in the top left of each panel. The coloured symbols
  represent the different simulation codes, as indicated on Fig. \ref{fig:md}.
  Again, the NR runs are separated into non-classic and classic SPH subgroups,
  while the FP runs are separated into AGN and noAGN subgroups, as indicated by
  the errorbars in the legends.}
\label{fig:cd}
\end{figure*}
%%%%%%%%%%%%%%%%%%%%%%%%%

The results so far suggest that there should be a measurable difference in
the concentration parameter between the different sets of runs. We investigate
this by assuming that the spherically averaged dark matter density profile,
$\rho(r)$, can be approximated by the \citet[][]{nfw1996, nfw1997} form,
\begin{equation}
  \label{eq:nfw}
  \frac{\rho(r)}{\rho_{crit}} = \frac{\delta_c}{ (r/r_s) (1+r/r_s)^2},
\end{equation}
here $\rho_{crit}$ is the critical density of the Universe, $\delta_c$ a
characteristic density, and $r_s$ a characteristic radius that is directly
related to the concentration $c_{NFW} = R_{200} /r_s$.

There is an extensive literature on the accuracy with which equation \ref{eq:nfw}
describes density profiles in dark matter only simulations, and while it
represents a reasonable approximation to the ensemble averaged density
profile of dark matter haloes in dynamical equilibrium, it cannot capture
the shape of the density profile in detail. The presence of baryons complicates
matters even further, as shown by \cite{Schaller2015a}, but equation
\ref{eq:nfw} provides a reasonable description of the dark matter density
profile over the radial range [$0.05R_{200} - R_{200}$].

Following \cite{Schaller2015a}, we fit both NFW parameters (i.e. $\delta_c$
and $r_s$) to the dark matter density profile within this radial range in the
DM, NR, and FP runs, using the {\small curve\_fit} package from {\small scipy}
\citep{scipy} with equally spaced logarithmic bins. In Fig. \ref{fig:dmfit},
we show residuals corresponding to these NFW fits using data drawn from the
classic and non-classic SPH examples, \gadgetmusic\ and \arepo; solid, dashed,
and dotted lines indicate DM, NR, and FP runs. Note that there are two
versions of the FP runs for each code. Within the fitting radius
range, which is indicated by the thick lines, the dark matter component mass
profile agrees with the NFW profile to within $\sim$ 15 \% (slightly
worse at $z$=1) for all three baryonic models.

In Fig. \ref{fig:cd}, we show how the ratio of concentration in the NR and FP
runs (left and right panels) relative to the DM run varies with measured
concentration. Within each panel, the left (right) section shows the $z$=0 (1)
trend. The behaviour in both the NR and FP runs is similar. At $z$=0,
the concentration is enhanced in both the classic and non-classic SPH runs,
and in the AGN and noAGN runs, to a similar extent, a factor of $\sim 1-1.2$.
At $z$=1, the enhancements are more pronounced in all of the NR and FP runs,
a factor of $\sim 1.5$, although the spread in values is larger in the
FP runs. Interestingly, for the NR runs at $z$=0, we see a clear separation in
the median value and enhancement of the concentration, with the classic SPH
runs showing a higher concentration and enhancement, consistent with our
observations in the previous section.

The concentration enhancements in the NR runs and the noAGN FP runs
are consistent with \cite{Duffy2010} and \citet{Fedeli2012}. The increased
concentration found in the FP runs with AGN feedback is in agreement with
\cite{Schaller2015a}, but contradicting \cite{Duffy2010}, who found either no
change or a decrement in concentration. We caution that our small number
statistics may play a role in the difference.

\subsection{Spin Parameter} %%, $\lambda$}
%%%%%%%%%% FIG 5 %%%%%%%%
\begin{figure*}
\includegraphics[width=1.0\textwidth]{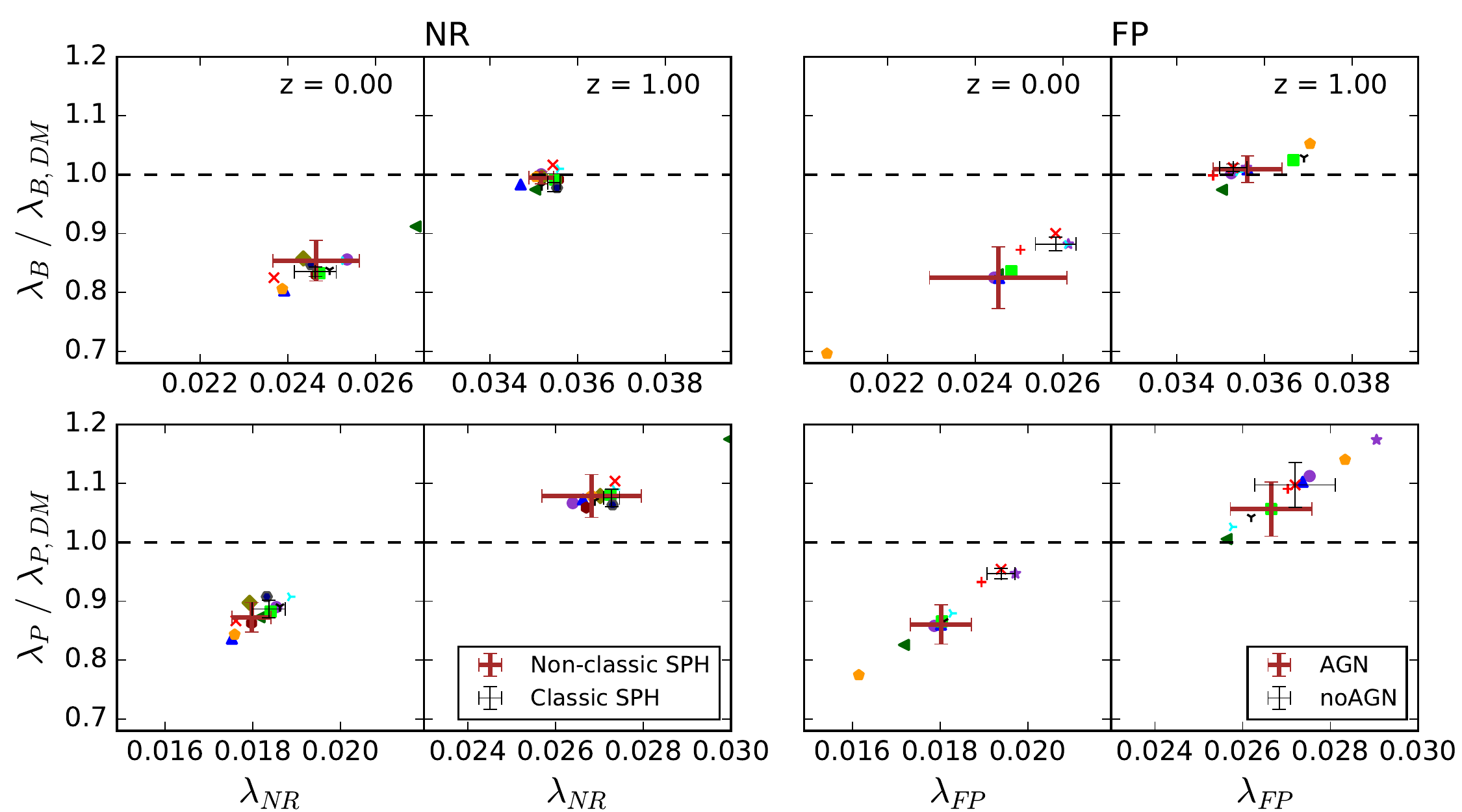}
\caption{Spin parameter changes with respect to the DM runs. The left column shows the
  results from NR runs, while the right column is for the FP runs. The two
  different methods (indicated as $\lambda_B$ \citep{Bullock2001}, $\lambda_P$
  \citep{Peebles1969}, in the y-label) are shown in each row. There are two
  subplots for each panel, which show the results at the two redshifts as
  indicated in the uppermost panels. The coloured symbols represent the different simulation
  codes, as indicated on Fig. \ref{fig:md}. Again, we separate the NR runs
  into non-classic SPH and classic SPH subgroups, the FP runs
  into AGN and noAGN subgroups, as indicated by the errorbars in the legends.}
\label{fig:sd}
\end{figure*}
%%%%%%%%%%%%%%%%%%%%%%%%%

The spin parameter $\lambda$ is commonly used to quantify the degree to which
the structure of a system is supported by angular momentum. Several definitions
for spin have been proposed, but we investigate the two most common definitions;

\smallskip

\noindent $\bullet$ \emph{$\lambda_P$}, the dimensionless ``classical'' spin
parameter \citep{Peebles1969},
\begin{equation}
  \lambda_P = \frac{J \sqrt{|E|}}{G M^{5/2}},
\end{equation}
where $J$ is the magnitude of the angular momentum of material within the virial
radius, $M$ is the virial mass, and $E$ is the total energy of the system; and

\smallskip

\noindent $\bullet$ \emph{$\lambda_B$}, the modified spin parameter of
\citet{Bullock2001}, which avoids the expensive calculation of the
total energy $E$ of a halo,
\begin{equation}
  \lambda_B = \frac{J}{\sqrt{2} M V R},
\end{equation}
here $V=\sqrt{GM/R}$ is the circular velocity at the virial radius $R$, and
$M$ and $J$ have the same meaning as in the ``classical'' spin parameter
$\lambda_P$. Both spin parameters are calculated including all material with
$r \le R_{200}$.

The spin parameters measured in the NR and FP runs are shown in the left
and right panels respectively of Fig. \ref{fig:sd}; coloured symbols are as
in Fig. \ref{fig:md}. FP runs are grouped into AGN (brown thick errorbars)
and noAGN models (black thin errorbars); NR runs are separated into
non-classic (brown thick errorbars) and classic SPH (black thin errorbars) runs.

There are a couple of points worthy of note in this Figure. Firstly, there is a
systematic drop between $z$=1 and $z$=0 in the ratio of $\lambda_B$ and
$\lambda_P$ with respect to their DM counterparts in both the NR and FP
runs and in all of the groupings (classic vs non-classic SPH, AGN vs noAGN).
Secondly, the measured spins are broadly similar in the NR runs, independent
of either redshift or classic vs non-classic SPH grouping, but there is a
much larger spread in values in the FP runs, and the result is sensitive to
whether or not AGN is included.

Interestingly, \cite{Bryan2013} found that the $z$=0 spin distribution of
dark matter haloes extracted from runs including baryonic physics, both with
and without AGN feedback, is not significantly different from that of dark
matter only haloes. They reported that their baryon runs exhibit slightly
lower median spin values at $z$=2 than in their dark-matter-only runs, in
apparent contradiction to our results. However, their median halo mass is
$M_{200} = 2 \times 10^{12} \hMsun$, which is about 3 orders lower than our
cluster, and these systems will have significantly different merging histories
than our cluster. Merging history is likely to influence the angular momentum
content of the system, especially that of the gaseous component, with angular
momentum cancellation occurring in response to collisions and shocks of gas
from multiple infall directions.

\subsection{Shape of Isodensity and Isopotential Shells}

%%%%%%%%%% FIG 8 & 9 %%%%%%%%
\begin{figure*}
\includegraphics[width=1.0\textwidth]{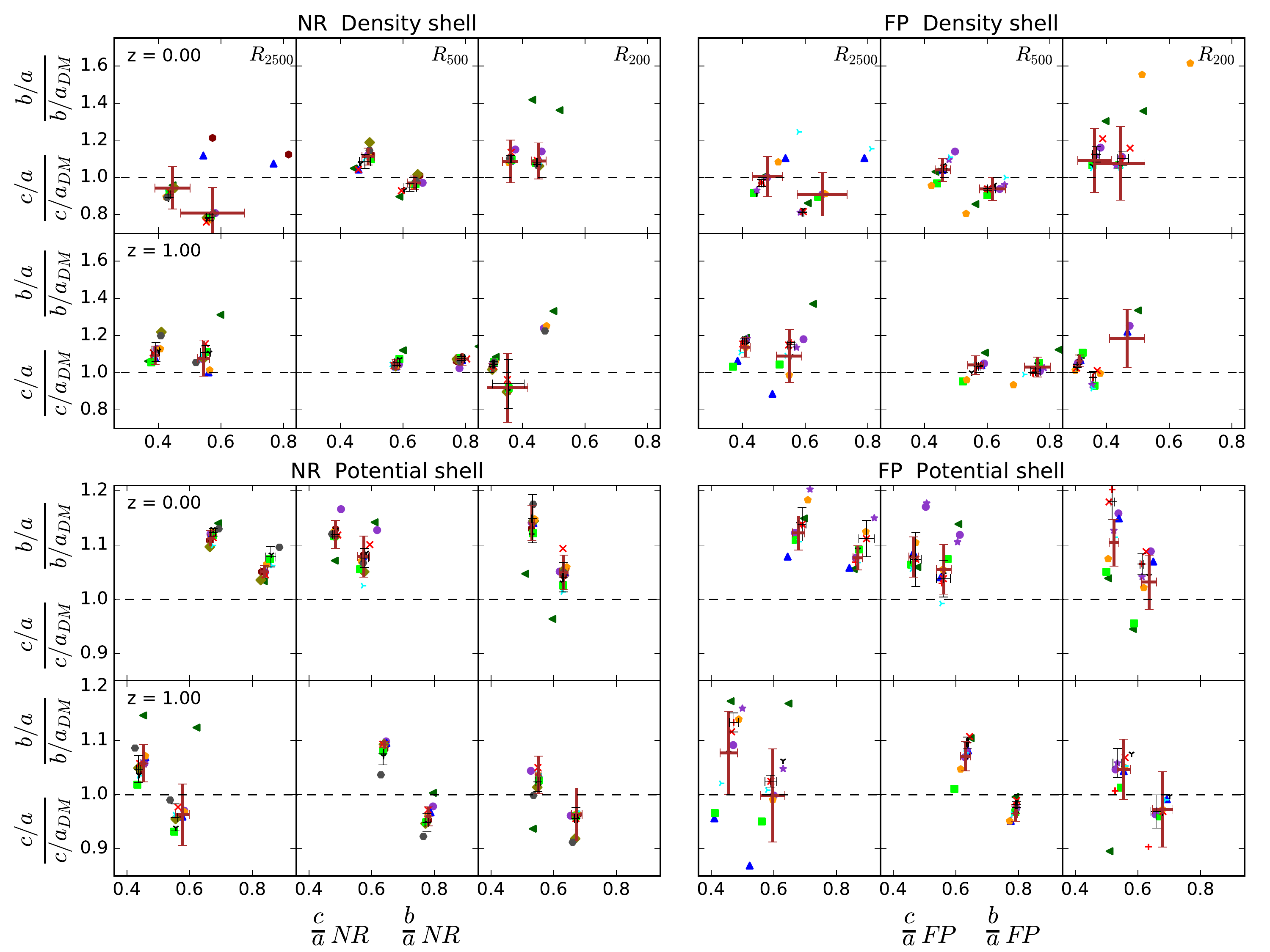}
\caption{The halo shape (axis ratios: $\frac{c}{a}$ and $\frac{b}{a}$) changes
  between the DM runs and NR runs (left column) / FP runs (right column) from
  both the isodensity shells (top two panels) and the isopotential shells (lower two
  panels). These results are calculated through the inertia method. We refer to
  Fig. \ref{fig:md} for the meanings of the coloured symbols. Inside each
  panel, we show the results at three shells at $R_{2500}, R_{500}$ and
  $R_{200}$ from left to right within each subplot, and at redshifts of $z$=0
  and $z$=1 in the top and bottom subplots. Again, the errorbars from the FP
  runs are grouped
  into AGN (brown thick errorbars) and noAGN (black thin errorbars); while the
  errorbar from the NR runs are grouped into non-classic SPH (brown thick
  errorbars) and classic SPH (black thin errorbars) methods.}
\label{fig:hsdd}
\end{figure*}

%\begin{figure*}
%\includegraphics[width=1.0\textwidth]{Tria_diff-eps-converted-to.pdf}
%\caption{Halo triaxiality changes inside density shell (upper panel) and
%  potential shell (lower panel) at the three radii based on the inertia method.
%  This plot is similar as Fig. \ref{fig:hsdd}, we refer to that figure for more
%  details.}
%\label{fig:tria}
%\end{figure*}
%%%%%%%%%%%%%%%%%%%%%%%%%

Having considered the spin parameter, we now move on to the shape of the
cluster's isodensity and isopotential surfaces. We adopt the common
method of diagonalization of the inertia tensor and characterization with
ellipsoids of either the interpolated density field \citep[e.g.][]{Jing2002} or
the underlying gravitational potential
\citep[e.g.][]{Springel2004,Hayashi2007,Warnick2008}. Following
\cite{Bett2007,Warnick2008}, the inertia tensor \citep[see][for more
discussions of the choice of inertia tensor]{Warnick2008,Vera-Ciro2011} is
defined as:
\begin{equation}
I_{\alpha\beta} = \sum^N_{i=1} m_i (r^2_i\delta_{\alpha\beta} - x_{i,\alpha}x_{i,\beta}),
\end{equation}
where $r_i$ is the position vector of the $i$th particle, $\alpha$ and $\beta$
are tensor indices ($\alpha,\beta=1,2,3$), $x_{i,\alpha}$ are components of the
position vector of $i$th particle, and $\delta_{\alpha\beta}$ is
the Kronecker delta. We estimate the shape of isodensity and isopotential shells
at three radii: $R_{2500}, R_{500}$ and $R_{200}$, selecting all particles (including dark matter, star and gas components) within
these shells as described in Appendix~\ref{A:is}. Eigenvalues can be computed by
noting that
\begin{equation}
I = \frac{M}{5}
  \begin{bmatrix}
  b^2+c^2 & 0 & 0 \\
  0 & a^2+c^2 & 0 \\
  0 & 0 & a^2+b^2
  \end{bmatrix}
\end{equation}
These axes then describe a hypothetical uniform ellipsoid whose axes $a \ge b
\ge c$ are those of the halo itself. Thus, we can have $b/a = \sqrt{(I_a + I_c -
  I_b) / (I_b + I_c - I_a)}$ and $c/a = \sqrt{(I_a + I_b - I_c) / (I_b + I_c -
  I_a)}$. For completeness, we also use a direct linear least squares fitting
method to
fit ellipsoids to the 3D isodensity surfaces to verify our results, which we
describe in Appendix \ref{A:fit}.
% It's useful to define a triaxiality parameter
% to interpret the significance of the computed axis ratios for the overall shape;
% this is
% \begin{equation}
% T = \frac{a^2-b^2}{a^2-c^2}.
% \end{equation}
% Low values of $T$ (i.e. $T \rightarrow 0$) correspond to oblate haloes, while
% high values (i.e. $T \rightarrow 1$) correspond to prolate haloes.

In Fig. \ref{fig:hsdd}, we show how the axis ratios, $b/a$ and $c/a$, change
between the DM runs and the corresponding
NR and FP runs (left and right columns) within thin isodensity and isopotential
shells (upper and lower panels) at $R_{2500}, R_{500}$ and $R_{200}$ (left,
middle, and right panels within each column) as a function of $b/a$ and $c/a$
in the NR and FP runs; the relevant redshift is shown in the leftmost panel of each row.

Broadly similar trends are evident in both the NR and FP runs at both
redshifts. At $z$=0, the isopotential shells become slightly rounder at all
radii, by a factor of $\sim 1.1-1.2$. The outermost isodensity shell becomes
slightly rounder by a similar factor; the inner shells become more oblate,
with negligible change in $c/a$, but $b/a$ drops by a factor of $\sim 0.8$.
At $z$=1, the trend is such that the inner isodensity shells become slightly
rounder by a factor of $\sim 1.1$ in both the NR and FP runs, whereas the
outermost shell can be either more oblate (NR) or prolate (FP). The isopotential
shells change in such a way that $c/a$ is enhanced whereas $b/a$ is reduced,
resulting in negligible net change in the overall shape of the halo.

\smallskip

The effect of including baryonic physics on the shapes of dark matter haloes
has been studied previously with hydrodynamic simulations in
\cite[][etc.]{Kazantzidis2004,Knebe2010,Bryan2013,Tenneti2014,Butsky2015,Velliscig2015}.
\cite{Kazantzidis2004} found that halos formed in simulations with gas cooling
are significantly more spherical than corresponding halos formed in adiabatic
simulations. \cite{Knebe2010} found that the inclusion of gas physics has no affect
on the (DM) shapes of subhaloes, but an influence on their suite of host haloes,
which drives the DM halo to become more spherical especially at the central regions
\citep[see also][etc.]{Debattista2008,Tissera2010,Abadi2010,Bryan2013,Tenneti2014,Butsky2015,Tenneti2015}.
Our results from the isopotential shell are in agreement with these literatures.
However, at the most inner isodensity shell -- $R_{2500}$,
there is a decrease of $b/a$ (slightly smaller decrease for $c/a$). However, the
increases for both $b/a$ and $c/a$ at $R_{2500}$ are very clear from the isopotential shell. This is possibly
caused by the substructures in the isodensity shell, which has less effect with the isopotential shell method.
Using hydro-dynamical simulations with different versions of baryon models, \cite{Velliscig2015}
showed these different baryon models have less effect on the halo shape. This agrees with our findings from
Fig. \ref{fig:hsdd}, which shows a broadly agreement between different simulation codes
as well as between the NR and FP runs.

\subsection{Velocity Anisotropy}

%%%%%%%%%% FIG 10 %%%%%%%%
\begin{figure*}
\includegraphics[width=1.0\textwidth]{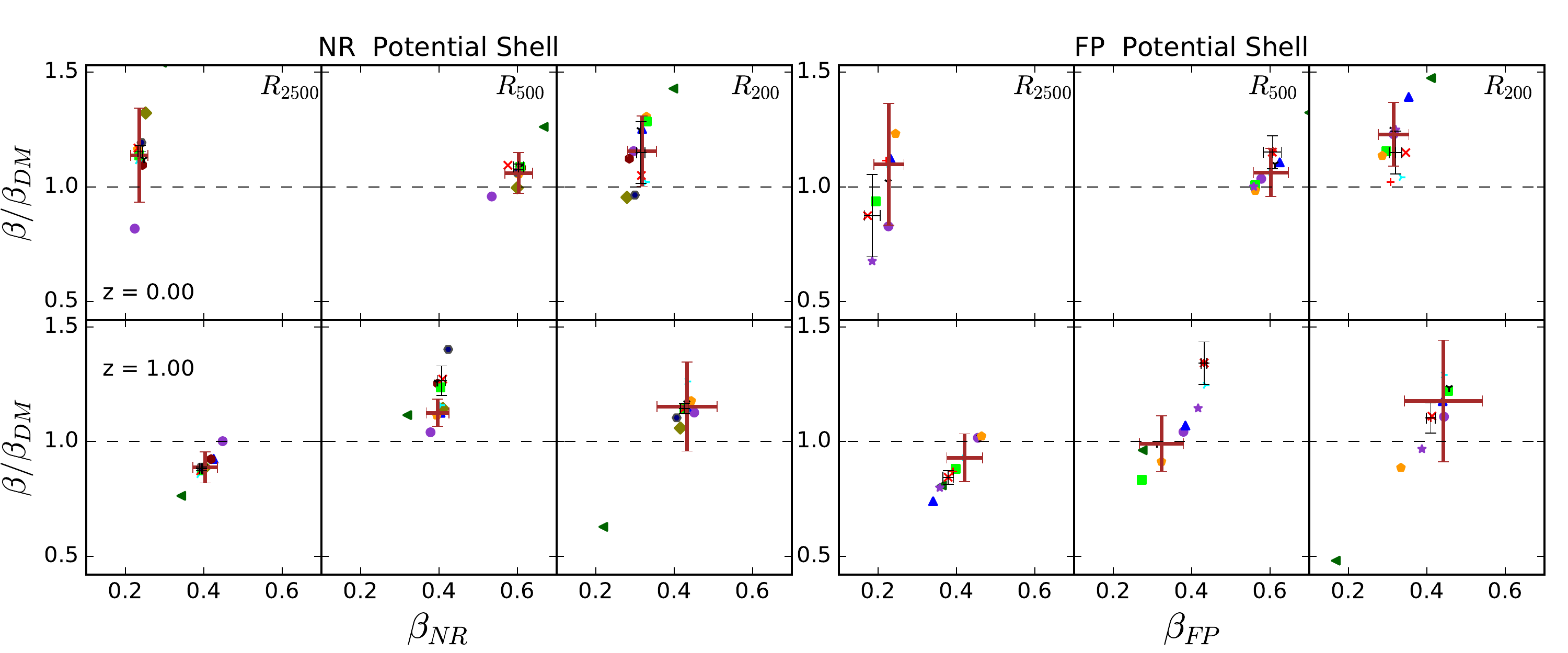}
\caption{The halo velocity anisotropy, $\beta$. The difference between the DM runs
  and NR runs (left column) / FP runs (right column). This figure is very
  similar to Fig. \ref{fig:hsdd} in subplot distribution, symbols and errorbars.
  We refer to Fig. \ref{fig:hsdd} for more details.}
\label{fig:had}
\end{figure*}
%%%%%%%%%%%%%%%%%%%%%%%%%

We finish our analysis by looking at the velocity anisotropy
\begin{equation}
  \beta = 1 - \frac{\sigma_{\rm tan}^2}{2 \sigma_{\rm r}^2},
\end{equation}
where $\sigma_{\rm tan}$ and $\sigma_{\rm r}$ are the tangential and radial
velocity dispersions. We compute these components of the velocity dispersion
using the particles selected in the isopotential shells at the three radii,
and show the results in Fig. \ref{fig:had}, revealing how $\beta$ varies between
NR and FP runs (left and right columns) at $z$=0 and 1 (upper and lower rows
within each column) at $R_{2500}, R_{500}$ and $R_{200}$ (left, middle, and
right panels within each column).

% At $z$=0, the velocity anisotropy is enhanced within the innermost shell,
% whereas a
Again, we see very similar values and changes of the $\beta$ parameter between the NR and FP
runs at fixed radius and redshift. At redshift $z$=1, we have larger $\beta$ values at $R_{200}$ and $R_{2500}$ than at $R_{500}$;
while at $z$=0 the $\beta$ value is much larger at $R_{500}$ than at the other two radii. The incrementation of $\beta$
at $R_{200}$ is $\sim$ 10 \% at both redshifts; at $R_{500}$, there is a slightly small
increase of $\beta$ ($\sim$ 5 \%) at $z$=0, while there are large disagreements between the subgroups at $z$=1;
at the innermost radius $R_{2500}$, there are about 10 \% increase of $\beta$ at $z$=0, but
about 10 \% decrease of $\beta$ at $z$=1 compared to their DM runs. Similar to the halo
shape changes, we do not find a clear separation between these subgroups, except the ones at $R_{500}$ and $z$=1.
There are also broad agreements between the results from the isodensity and from the isopotential shells.

\smallskip

\cite{Lau2010} investigated two hydro-dynamical simulations: one with no-radiative gas;
the other including gas cooling, star formation and feedback. By comparing the two,
they found that the dark matter velocity anisotropy profile is almost
unaffected by the addition of cooling, star formation and feedback and insensitive to
redshift between $z$=0 and 1. This is in very good agreement with what we find in Fig.
\ref{fig:had} -- there are very similar values and changes of the $\beta$ parameter between the NR and FP runs.

%%%%%%%%%%%%%%%%%%%%%%%%%%%%%%%%%%%%%%%%%%%%%%%%%%%
\section{Discussion and Conclusions} \label{sec:conclusions}
%%%%%%%%%%%%%%%%%%%%%%%%%%%%%%%%%%%%%%%%%%%%%%%%%%%

We have investigated the performance of 11 modern astrophysical simulation
codes -- \hydra, \arepo, \ramses\ and 8 versions of \gadget\ with different SPH
implementations -- and with different baryonic models -- by carrying out
cosmological zoom simulations of a single massive galaxy cluster. By comparing
different simulation codes and different runs ranging from dark matter only
to full physics runs, which incorporate cooling, star formation, black hole
growth, and various forms of feedback, we set out to
\begin{itemize}
  \item[(i)] {\it quantify the scatter between codes and different baryon models.}
  \item[(ii)]{\it understand the impact of baryons on cluster properties, and the
    extent to which these properties converge.}
\end{itemize}
For clarity, and motivated by the results of \citetalias{Sembolini2015}, we
grouped codes according to whether or not they are ``Classic SPH'', which
produce declining entropy profiles with decreasing radius in non-radiative
runs, or ``non-Classic SPH'', which include the mesh, moving mesh, and modern
SPH codes, which recover entropy cores at small radii. We also grouped full
physics runs according to whether or not they include black hole growth and
AGN feedback as ``AGN'' and ``noAGN'' runs respectively. Our key findings can
be summarised as follows:

\begin{itemize}

  \item[]{\bf Code Scatter:} In \citetalias{Sembolini2015}, we
    already saw that code-to-code scatter between codes for the
    aligned dark-matter-only runs is within 5 \% for the total
    mass profile. If we ignore this difference, the non-radiative gas boosts
    this scatter up to $\sim 30$ \% at $z$=0, with the largest difference
    evident in the central regions, and up to $\sim 50$ \% at
    $z$=1. However, by grouping codes into classic and non-classic SPH,
    the scatter for the total mass profile within a grouping is reduced to
    $\sim$ 20 \%; this means that the disagreement is driven by
    the approach to solving the equations of gas dynamics.
    The scatter for the total density
    profile is reduced to $\le 5$ \% between all codes at $R
    \ge R_{2500}$, and even smaller at larger radius.

    The scatter in the total mass profile between different codes in
    the FP runs, when compared to the NR runs, is larger -- over 100 per
    cent at $z$=0, greater at $z$=1, within the central regions.
    Grouping the runs into those that include AGN feedback and those that do
    not, the scatter in the central regions is still substantial, which
    implies that the complexities of sub-grid physics can produce
    very different results, even when similar baryonic physics prescriptions
    are adopted. This is especially true for the codes with AGN feedback. The
    scatter between different runs reduces to $\le 10$ per
    cent at $R \approx R_{2500}$, and smaller at
    larger radii.

    {\it For most of the global cluster properties investigated in this paper, we
      find the scatter between different codes and different baryonic physics
      models is within $\sim$ 20 \%, in agreement with \citetalias{Sembolini2015,Sembolini2015b,Elahi2015}.}

  \item[]{\bf Impact of Baryons:} Using the DM runs as our reference, we find
    that the change in total mass profile in the FP runs is more marked than in
    the NR runs, especially within the central regions. Already within $R_{500}$
    we see $\sim 10$ \% variations with respect to the median in the
    FP runs, which grows to $\sim 20$ \% variations at $R_{2500}$.
    In contrast, the variations with respect to the median are markedly
    smaller in the NR runs, $\simlt 10$ \% at $R_{2500}$. The impact
    on the central density appears to be redshift-, code-, and
    physics-dependent, insofar as we see a largely uniform trend for
    lower central densities in the NR runs at $z$=0; enhanced central
    densities in the FP runs at $z$=1; and a mixture of behaviours in the NR
    and FP runs at $z$=1 and 0 respectively, although it is noteworthy that
    it is the non-classic SPH and AGN that produce lower central densities,
    as we might expect. Overall, we conclude that the scatter between the codes
    in the NR runs is less important than the scatter between different
    baryonic physics models in the FP runs.

    {\it Although the different global cluster properties have different
    responses to baryon physics, there is broad agreement at both
    redshifts between the NR and FP runs,and with the conclusions of
    \citetalias{Sembolini2015b}.} Because of the large scatter of the
    total mass profile in the central regions, the total inner density
    slope $\gamma$ and the concentration $C_{NFW}$, shows the largest scatter,
    with a clear separation between the different subgroups at $z$=0.

    By choosing the three characteristic radii -- $R_{2500}, R_{500}$ and
    $R_{200}$, we investigate how the cluster properties change at different
    radii. The halo mass changes have a clear radius dependence at both
    redshifts, the inner radius shows the largest increase for both the NR and
    FP runs compared to the DM runs. There is almost no mass change for
    $M_{200}$ at both redshifts. The halo shape changes are dependent on
    the choice of the shells; isodensity shells change from inner to
    outer radii, but are weakly dependent on redshift, whereas isopotential
    shell changes are systematic with radius and redshift.
\end{itemize}

%about the AGN - noAGN
It's interesting to note that the clear separation we see between
classic and non-classic SPH runs in the mass profiles in the NR runs is not
reproduced in the FP runs. How much of this difference is driven by the
hydrodynamical technique?
In the AGN runs (right upper panel of Fig. \ref{fig:adp}), the
classic SPH codes \gadgettwox\ and \gadgetowls\ tend to have much higher density
at the cluster centre than the non-classic SPH codes \gadgetx, \arepo\ and
\ramses, while the non-classic SPH codes \gadgetpesph, which uses a heuristic
model to quench star formation, produces a much lower density
profile than the other codes from the noAGN group. In addition, the gas profile
difference between these simulation codes in the NR runs is about 100 \%
at the cluster centre, as was shown in \citetalias{Sembolini2015}. This seems to suggest
that the hydrodynamic technique can be as important as baryonic physics
in setting the mass profile in the FP runs. However, we note also that the
total mass profile in the non-classic SPH code -- \areposh\ -- that does not
include AGN feedback is very close to the classic SPH codes without AGN
feedback, and the non-classic SPH code \gadgetmagneticum\ has a higher
central density than codes that do not include AGN feedback, despite having
AGN feedback included. This suggests that the hydrodynamic scheme may be
important, but the details of the baryonic physics prescription is more
important in shaping the mass profile.

There are two FP runs of \gadgetmusic\ in this study, the original one runs with \gadgetthree\ code and the \cite{Springel2002} baryon model; while the other one -- \gadgetmusicpi\ run with \gadget\ code and the \cite{Piontek2011} baryon model. Through this study, we find that there is almost no difference between the two simulations, which can be understood as there are no differences between the two simulation codes and between the two versions of baryon models for this simulated galaxy cluster.

Although we have shown the scatter between different simulation codes / techniques and between different baryonic models, a detailed comparison of the algorithms as well as of the numerical implementation methods of baryonic models is in great needs, because these details are essential for explaining the scatter we show in this paper. To achieve this goal, we are planning to first perform a convergence test in a following study, and then extend this comparison project to an extensive examination on these parameters in the baryon models.

Although this work is based on the analysis of only one simulated galaxy cluster, we argue that our results are robust, because most of them are mainly shown by the differences, in which most systematic errors should be canceled. However, it will be worth to increase the statistics by simulating more clusters in further comparisons: for example, relaxed and un-relaxed clusters may give different answers due to their different dynamical state. We are including more MUSIC clusters to our comparison project and will present the results in future papers.

% The shallower mass profile for NR runs at the cluster centre region agrees
% with the findings of e.g. \cite{Cui2012a}. This could be caused by
% stripped gas as the substructures falling into the centre of the cluster. This
% gas component, unlike dark matter particles, is uneasily merging into the
% cluster centre because of the hydro-dynamic viscosity. While the increasing
% mass profile at cluster centre region for the FP runs are in agreement with
% e.g. \cite{Cui2014a, Velliscig2014}.

% In the recent Aquila comparison project, \cite{Scannapieco2012} compared
% the results of various cosmological gas-dynamical codes used to simulate the
% formation of a galaxy, and claimed that despite the large spread in properties
% spanned by the simulated galaxies, none of them has properties fully consistent
% with theoretical expectations or observational constraints. State-of-the-art
% simulations cannot yet uniquely predict the baryonic
% properties of a galaxy, even when the assembly history of its host halo
% is fully specified. We will compare the full physics models with observations in
% the follow-up work.

\section*{Acknowledgements}

The authors would like to thank Joop Schaye and Stefano Borgani for their kind helps and suggestions.
The authors would like to acknowledge the support of the International Centre
for Radio Astronomy Research (ICRAR) node at the University of Western Australia
(UWA) for the hosting of the ``Perth Simulated Cluster Comparison" workshop in
March 2015, the results of which has led to this work; the financial support of
the UWA Research Collaboration Award (RCA) 2014 and 2015 schemes; the financial
support of the Australian Research Council (ARC) Centre of Excellence for All
Sky Astrophysics (CAASTRO) CE110001020; and ARC Discovery Projects DP130100117
and DP140100198. We would also like to thank the Instituto de Fisica Teorica
(IFT-UAM/CSIC in Madrid) for its support, via the Centro de Excelencia Severo
Ochoa Program under Grant No. SEV-2012-0249, during the three week workshop
``nIFTy Cosmology" in 2014, where the foundation for much of this work was
established.\\

\noindent WC acknowledges support from UWA RCAs PG12105017 and PG12105026, and
from the Survey Simulation Pipeline (SSimPL; {\texttt{http://www.ssimpl.org/}}).

\noindent CP is supported by an ARC Future Fellowship FT130100041 and ARC
Discovery Projects DP130100117 and DP140100198.

\noindent AK is supported by the {\it Ministerio de Econom\'ia y Competitividad} (MINECO)
in Spain through grant AYA2012-31101 as well as the Consolider-Ingenio 2010
Programme of the {\it Spanish Ministerio de Ciencia e Innovaci\'on} (MICINN)
under grant MultiDark CSD2009-00064. He also acknowledges support from ARC
Discovery Projects DP130100117 and DP140100198. He further thanks Dylan
Mondegreen for something to dream on.

\noindent PJE is supported by the SSimPL programme and the Sydney Institute for
Astronomy (SIfA), and {\it Australian Research Council} (ARC) grants DP130100117
and DP140100198.

\noindent GY and FS acknowledge support from MINECO (Spain) through the grant
AYA 2012-31101. GY thanks also the Red Espa\~{n}ola de Supercomputacion for
granting the computing time in the Marenostrum Supercomputer at BSC,
where all the MUSIC simulations have been performed.

GM acknowledges supports from the PRIN-MIUR
2012 Grant "The Evolution of Cosmic Baryons" funded by the
Italian Minister of University and Research, from the PRIN-INAF
2012 Grant "Multi-scale Simulations of Cosmic Structures" funded by the Consorzio per la Fisica di Trieste.

\noindent AMB is supported by the DFG Cluster of Excellence "Universe" and
by the DFG Research Unit 1254 "Magnetisation of interstellar and intergalactic media".

\noindent CDV acknowledges support from the Spanish Ministry of Economy and Competitiveness (MINECO) through grants AYA2013-46886 and AYA2014-58308. CDV also acknowledges financial support from MINECO under the 2011 Severo Ochoa Program MINECO SEV-2011-0187.

\noindent EP acknowledges support by the Kavli foundation and the ERC grant
"The Emergence of Structure during the epoch of Reionization".

\noindent RJT acknowledges support via a Discovery Grant from NSERC
and the Canada Research Chairs program. Simulations were run on
the CFI-NSRIT funded Saint Mary's Computational Astrophysics
Laboratory.\\

\noindent The authors contributed to this paper in the following ways: WC, CP,
\& AK
formed the core team that organized and analyzed the data, made the plots and
wrote the paper. CP, WC, LO, AK, MK, FRP \& GY organized the nIFTy workshop at
which this program was completed. GY supplied the initial conditions. PJE
assisted with the analysis. All the other authors, as listed in Section 2
performed the simulations using their codes. All authors %had the opportunity to proof
read and comment on the paper.\\

\noindent The simulation used for this paper has been run on Marenostrum
supercomputer and is publicly available at the MUSIC website.

\noindent The \arepo\ simulations were performed with resources awarded through
STFCs DiRAC initiative.

\noindent \gadgetsphs\ sumulations were carried out using resources provided by the
Pawsey Supercomputing Centre with funding from the Australian Government and the
Government of Western Australia.

\noindent \gadgetpesph\ simulations were carried out using resources at the Center
for High Performance Computing in Cape Town, South Africa.

\noindent \gadgetanarchy\ simulations were performed on the Teide High-Performance
Computing facilities provided by the Instituto Tecnol\'{o}gico y de Energ\'{i}as Renovables (ITER, SA).

\noindent This research has made use of NASA's Astrophysics Data System (ADS)
and the arXiv preprint server.

\noindent All the figures in this paper are plotted using the python matplotlib
package \citep{Hunter:2007}.

%*****************************************************************************
\bibliographystyle{mnras}
\bibliography{bibliography}

\appendix

\section{Simulation codes} \label{A:codes}

\noindent \paragraph*{RAMSES} (Perret, Teyssier)

\noindent \ramses\ is based on adaptive mesh refinement (AMR) technique, with a
tree-based data structure allowing recursive grid refinements on a cell-by-cell
basis. The hydrodynamical solver is based on a second-order Godunov
method with the HLLC Riemann solver. For the baryon physics, \ramses\ modifies
\cite{Haardt1996} for the gas cooling and heating with metal cooling function of
\cite{sutherland_dopita93}. The UV background and a self-shielding recipe is
based on \cite{Aubert2010}. The star formation follows \cite{Rasera2006} with
density threshold of $n_* = 0.1 H cm^{-3}$. The formation of super massive black
hole (SMBH) uses the sink particle technique \citep{Teyssier2011}.
The SMBH accretion rate can have a boost factor compared to the Bondi accretion
rate \citep{BoothSchaye2009}. It can not exceed the instantaneous Eddington
limit, however. The AGN feedback used
is a simple thermal energy dump with $0.1c^2$ of specific energy, multiplied by
the instantaneous SMBH accretion rate.

\noindent \paragraph*{AREPO} (Puchwein)

\noindent \arepo\ employs a TreePM gravity solver and the hydrodynamic equations
are solved with a finite-volume Godunov scheme on an unstructured moving Voronoi
mesh \citep{Springel10}. Detailed descriptions of the galaxy formation models
implemented in \arepoil\ can be found in \cite{Vogelsberger2013,
  Vogelsberger2014}. The other FP version (\areposh) of \arepo\ has the same
  baryon model as \gadgetmusic.

\noindent \paragraph*{G2-ANARCHY} (Dalla Vecchia)

\noindent \gadgetanarchy\ is an implementation of \gadget-2 \citep{Gadget2}
employing the pressure-entropy SPH formulation derived by
\cite{Hopkins13}. \gadgetanarchy\ uses a purely
numerical switch for entropy diffusion similar to the one of \cite{Price2008},
but without requiring any diffusion limiter. The kernel adopted is the $C^2$
function of \cite{Wendland95} with 100 neighbors, with the purpose of avoiding
particle pairing \citep[as suggested by][]{Dehnen2012}. A FP version of this
code is not available yet.

\noindent \paragraph*{G3-X} (Murante, Borgani, Beck)

\noindent Based on \gadgetthree, an updated version of \gadget, \gadgetx\
\citep{Beck2016} employs a Wendland $C^4$ kernel with 200 neighbours
\citep[cf.][]{Dehnen2012}, artificial conductivity to promote fluid mixing
following \cite{Price2008} and \cite{Tricco13}, but with an additional limiter
for gravitationally induced pressure gradients. In the FP run of \gadgetx, gas
cooling is computed for an optically thin gas and takes
into account the contribution of metals \citep{Wiersma2009a}, with a uniform UV
background \citep{Haardt2001}. Star formation and chemical evolution are implemented as in
\cite{tornatore_etal07}. Supernova feedback is therefore modeled as kinetic and
the prescription of \cite{Springel2003} is followed. AGN feedback follows the
model described in \cite{Steinborn2015}. It sums up both the AGN mechanical and
radiative power, which is a function of the SMBH mass and the accretion
rate \citep{Churazov2005} and gives the resulting energy to the surrounding gas,
in form of purely thermal energy.

\noindent \paragraph*{G3-SPHS} (Power, Read, Hobbs)

\noindent \gadgetsphs\ is a modification of the standard \gadgetthree\ code,
developed to overcome the inability of classic SPH to resolve
instabilities. \gadgetsphs\ uses as an alternative either the HOCT kernel with
442 neighbours or the Wendland C4 kernel with 200 neighbours, based on a higher
order dissipation switch detector. A FP version of this code is under
development.

\noindent
\paragraph*{G3-MAGNETICUM} (Saro)
%%%%%%%%%%%%%%%%%%%%%%%%%%%%%%%%%%%%%%%%%%%%%%%%%%%

\noindent \gadgetmagneticum\ is an advanced version of \gadgetthree. In the
non-radiative version, a higher order kernel based on the bias-corrected,
sixth-order Wendland kernel \citep{Dehnen2012} with 200 neighbors is included.
It also includes a low viscosity scheme to track turbulence \citep{Dolag2005,
Beck2016} and isotropic thermal conduction with $1/20^{th}$ Spitzer rate
\citep{Dolag2004}. For its FP runs, the simulation allows for radiative
cooling according to \cite{Wiersma2009a} with metal line cooling from the
\textsc{CLOUDY}
photoionization code \citep{Ferland1998}, and heating from a uniform
time-dependent ultraviolet background \citep{Haardt2001}. The star formation
model is improved from \cite{Springel2003}, and it also includes chemical
evolution model according to \cite{tornatore_etal07}. The stellar feedback
triggers galactic winds with a velocity of $350$ km/s. The detailed SMBH growth
and AGN feedback model is presented in \cite{Magn14,Dolag2015}.

\noindent
\paragraph*{G3-PESPH} (February, Dav\'e, Katz, Huang)
%%%%%%%%%%%%%%%%%%%%%%%%%%%%%%%%%%%%%%%%%%%%%%%%%%%

\noindent \gadgetpesph\ is an implementation of \gadgetthree\ with
pressure-entropy
SPH \citep{Hopkins13} which features special galactic wind models. The SPH
kernel is an HOCTS (n=5) B-spline with 128 neighbors. For the FP run, the
radiative cooling in this simulation code is described in \cite{Katz1996}, with
metal lines cooling \citep{Wiersma2009a} and a uniform ionizing UV background
\citep{Haardt2001}. The star formation in this code is based on
\cite{Springel2003}. In addition, the heuristic model of
\cite{Rafieferantsoa2015}, tuned to reproduce the exponential truncation of the
stellar mass function, is used to quench star formation in massive galaxies. It
uses a highly constrained heuristic model for galactic outflows, described in
detail in \cite{Dave2013}, which utilises outflows scalings expected for
momentum-driven winds in sizable galaxies, and energy-driven scalings in dwarf
galaxies. It does not include AGN feedback in this process.

\noindent
\paragraph*{G3-MUSIC} (Yepes)
%%%%%%%%%%%%%%%%%%%%%%%%%%%%%%%%%%%%%%%%%%%%%%%%%%%

\noindent The original MUSIC runs (\gadgetmusic) were done with the
\gadgetthree\ code, based on the entropy-conserving formulation of SPH
\citep{Springel2002}. \gadgetthree\ employs a spline kernel \citep{Monaghan1985}
and parametrize artificial viscosity following the model described by
\cite{Monaghan1997}. \gadgetmusic\ uses the basic \cite{Springel2003} model
without SMBH growth and AGN feedback for its FP runs. In addition, an
alternative version of MUSIC performed using the radiative feedbacks described
in \cite{Piontek2011} is presented as \gadgetmusicpi, which also does not
include SMBH growth and AGN feedback.

\noindent
\paragraph*{G3-OWLS} (McCarthy, Schaye)
%%%%%%%%%%%%%%%%%%%%%%%%%%%%%%%%%%%%%%%%%%%%%%%%%%%

\noindent The \gadgetowls\ is based on the TreePM-SPH code \gadgetthree, and
uses standard entropy-conserving SPH with 48 neighbours for
its NR runs. For the FP runs, it includes new sub-grid physics for radiative
cooling, star formation, stellar feedback, black hole growth and AGN feedback
\citep[see more details in][]{Schaye2004, Schaye2008, Dallavecchia2008,
Wiersma2009a, BoothSchaye2009, Schaye10}, also for stellar evolution and mass loss \citep[see more details in][]{Wiersma2009b}, which is
developed as part of the OWLS/cosmo-OWLS projects \citep{Schaye10,LeBrun2014}.

\noindent
\paragraph*{G2-X} (Kay)
%%%%%%%%%%%%%%%%%%%%%%%%%%%%%%%%%%%%%%%%%%%%%%%%%%%

\noindent \gadgettwox\ is a modified version of the \gadget-2\ code
\citep{Gadget2},
using the TreePM gravity solver and standard entropy-conserving SPH with 50
neighbours for its NR runs. More details of the baryon model for its FP runs can
be found at \cite{Pike2014}. Cooling follows the prescription of
\cite{Thomas1992}. Gas is converted to stars at a rate given by the
Kennicutt-Schmidt relation \citep{Kennicutt1998}. Star formation follows the
method of \cite{Schaye2008}. A prompt thermal Type II SNe feedback model is
used. The AGN feedback uses the \cite{BoothSchaye2009} model with
a variation.

\noindent \paragraph*{HYDRA} (Thacker)

\noindent HYDRA-OMP \citep[\hydra][]{Thacker06}, a parallel implementation of
the \hydra\ code of \citet{Couchman95}, adopts a ``classic'' SPH implementation
with 52 neigbours, standard pair-wise artificial viscosity, and conservative
time-stepping scheme that keeps all particles on the same synchronisation. No FP
runs is performed in this code.

\section{Halo Shape: density and potential shells} \label{A:is}

Both the density and potential shells at $R_{2500}, R_{500}$ and $R_{200}$ are
used to determine the halo shape through the inertia method. Here we describe
how we select out the density and potential shells consistently
\citep[similar to ][]{Warnick2008}:
\begin{enumerate}
  \item[1] The median density $\rho_x$ and potential $\Phi_x$ for the shell at
    the three radii are calculated from all particles within $0.95 \times R_x
    \le r \le 1.05 \times R_x$, where $x$ indicates the overdensity in
    $[2500, 500, 200]$. We have checked the median density and potential
    between different simulation codes, and find the differences are within
    15 \%.
  \item[2] All particles within the density shell ([$0.95 \times \rho_x, 1.05
    \times \rho_x$]) or potential shell ([$0.99 \times \Phi_x, 1.01
    \times \Phi_x$]) are selected. A smaller range is used for the
    potential shell, because the potential is much smoother than the density.
    We adopt a range in density and potential that is twice as large at $z$=1,
    to ensure that we have a sufficient number of particles to get a reliable
    estimate of shape.
  \item[3] All the selected particles are grouped by a FoF method with a large
    linking length of 50 $\kpc$. We use an even larger linking length of 100
    $\kpc$ for shells at $R_{200}$, because the particles at this radius have
    larger separation. The most massive FoF group is chosen. This procedure
    allows us to remove particles that are too far away from the shell. We have
    checked the number fraction of the most massive group, which is always
    larger than 80 \% of the total selected shell particles.
\end{enumerate}

It is well known that the reliability of shape estimates of particle
distributions depends on the number of particles used to trace those
distributions \citep[e.g.][]{Tenneti2014}. We have checked the total number of
particles selected from these shells, and confirm that none have fewer than
6000 particles.

\section{Halo Shape: Direct Linear Least Squares Fitting Method} \label{A:fit}

\begin{figure*}
\includegraphics[width=1.0\textwidth]{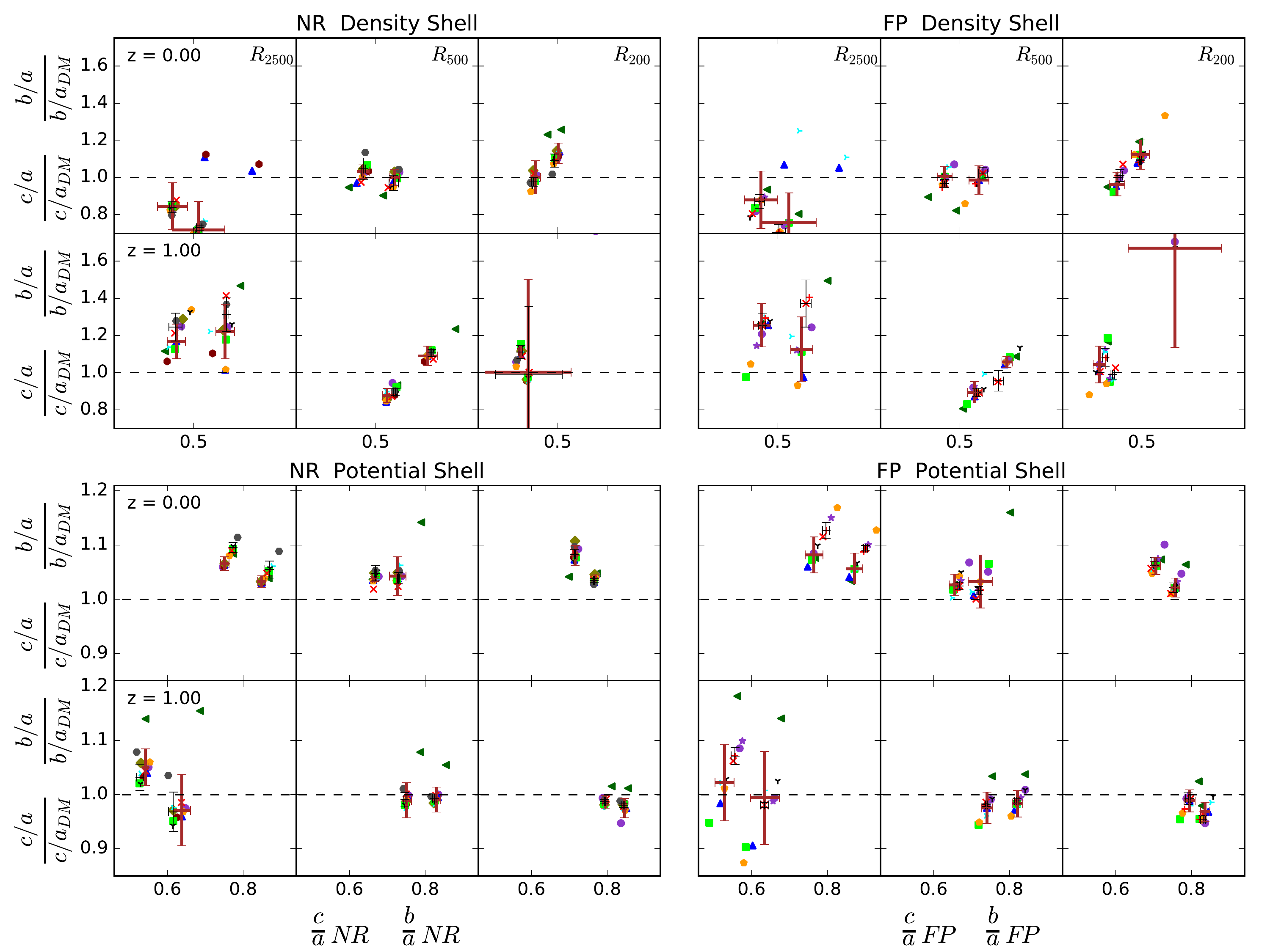}
\caption{Similarly to Fig. \ref{fig:hsdd}, but for the results from
  the fitting method. Refer to Fig. \ref{fig:hsdd} for the details.}
\label{fig:hspd}
\end{figure*}

To investigate the sensitivity of our results on how the halo shape changes
between the DM, NR and FP runs, we recompute halo shapes using a different
method, based on a direct linear least squares fit \footnote{Details
  of the fitting procedure can be found at here:
  http://www.mathworks.com/matlabcentral/fileexchange/24693-ellipsoid-fit},
to fit ellipsoids to the 3D isodensity surfaces. This fitting method uses the
same particles from both density and potential shells for the
inertia method. We note here that all particles inside the shell are treated
equally during the fitting, i.e. there is no mass weighting. With this fitting
method, we can directly estimate the length of the three axes: $a, b$ and $c$.

In Fig. \ref{fig:hspd}, we show how the halo shape changes ($b/a$ and $c/a$) in
the NR runs (left column) and FP runs (right column), focusing on three shells
corresponding to $R_{2500}, R_{500}$ and $R_{200}$ (from left to right) as a
function of $b/a$ and $c/a$; results for redshift $z$=0 (1) are shown in the
top (bottom) panels. The top (bottom) two rows show results for the density
(potential) shells. We refer to Fig. \ref{fig:md} for the meanings of the
colour symbols and errorbars. In each panel, we show both $c/a$ and $b/a$ data.

The values of $b/a$, $c/a$, and their changes with respect to the DM runs are
similar to the results in the corresponding panels in Fig. \ref{fig:hsdd}.
However, it is worth to note that for the isopotential shell, the fitting method
gives slightly smaller changes of both $b/a$ and $c/a$ at $z$=0; there is almost
no change of both $b/a$ and $c/a$ at $z$=1 and at $R_{500, 200}$; we also expect
there are slight variations in the size of the errorbars.
Besides that, We expect both methods are robust and precise for estimating the halo shape.

\bsp
\label{lastpage}
\end{document}